\documentclass[preprint,notoc]{JHEP3}
\usepackage{epsfig}

\def\eslt{\not\!\!{E_T}}
\def\eslt{E_T^{\rm miss}}

\def\to{\rightarrow}

\def\bi{\begin{itemize}}
\def\ei{\end{itemize}}
\def\te{\tilde e}
\def\th{\tilde h}
\def\tH{\tilde H}

\def\tu{\tilde u}

\def\tb{\tilde b}

\def\tH{\tilde H}
\def\tst{\tilde t}
\def\ttau{\tilde \tau}

\def\tg{\tilde g}

\def\tell{\tilde\ell}
\def\tq{\tilde q}

\def\tw{\widetilde W}
\def\tz{\widetilde Z}
\def\alt{\stackrel{<}{\sim}}
\def\agt{\stackrel{>}{\sim}}
\def\be{\begin{equation}}  
\def\ee{\end{equation}}  

\title{Mixed Higgsino Dark Matter\\ from a Reduced $SU(3)$ Gaugino Mass:\\
Consequences for Dark Matter and Collider Searches}
\author{Howard Baer$^a$, Azar Mustafayev$^a$, Eun-Kyung Park$^a$, 
Stefano Profumo$^{b}$ and \qquad \qquad Xerxes Tata$^c$\\
$^a$Department of Physics, Florida State University Tallahassee, 
FL 32306, USA\\
$^b$Division of Physics, Mathemathics and Astronomy, 
California Institute of Technology, Mail Code 106-38, Pasadena, CA 91125, USA\\
$^c$Department of Physics and Astronomy, University of Hawaii,
Honolulu, HI 96822, USA\\
E-mail: \email{baer@hep.fsu.edu},\email{mazar@hep.fsu.edu},\email{epark@hep.fsu.edu},
\email{profumo@caltech.edu},\email{tata@phys.hawaii.edu}}

\preprint{\vbox{FSU-HEP-060320, UH-511-1082-06}}

\abstract{ In gravity-mediated SUSY breaking models with non-universal
gaugino masses, lowering the $SU(3)$ gaugino mass $|M_3|$ leads to a
reduction in the squark and gluino masses. Lower third generation squark
masses, in turn, diminish the effect of a large top quark Yukawa
coupling in the running of the higgs mass parameter $m_{H_u}^2$, leading
to a reduction in the magnitude of the superpotential $\mu$ parameter
(relative to $M_1$ and $M_2$). A low $|\mu |$ parameter gives rise to
mixed higgsino dark matter (MHDM), which can efficiently annihilate in
the early universe to give a dark matter relic density in accord with
WMAP measurements.  We explore the phenomenology of the low $|M_3|$
scenario, and find for the case of MHDM increased rates for direct and
indirect detection of neutralino dark matter relative to the mSUGRA
model.  The sparticle mass spectrum is characterized by relatively light
gluinos, frequently with $m_{\tg}\ll m_{\tq}$.  If scalar masses are
large, then gluinos can be very light, with $\tg\to\tz_i g$ loop decays
dominating the gluino branching fraction.  Top squarks can be much
lighter than sbottom and first/second generation squarks.  The presence
of low mass higgsino-like charginos and neutralinos is expected at the
CERN LHC.
The small $m_{\tz_2}-m_{\tz_1}$ mass gap
should give rise to a visible opposite-sign/same flavor dilepton mass
edge.  At a TeV scale linear $e^+e^-$ collider, the region of MHDM will
mean that the entire spectrum of charginos and neutralinos are amongst
the lightest sparticles, and are most likely to be produced at
observable rates, allowing for a complete reconstruction of the
gaugino-higgsino sector.  }


\begin{document}

\section{Introduction}
\label{sec:intro}
 
Recently, a variety of astrophysical measurements 
by the WMAP\cite{wmap} and other collaborations have 
determined the density of cold dark matter (CDM) in the universe to be
\be
\Omega_{CDM}h^2=0.113\pm 0.009 .
\label{eq:wmap}
\ee
The additional determination of a non-zero dark energy 
component to the universe suggests that we live in a 
$\Lambda CDM$ universe, with $\Omega_\Lambda h^2\sim 0.35$.
While the nature of dark energy remains a mystery,
there are a number of well-motivated particle physics candidates for the
CDM,
and collider and DM search experiments may serve to distinguish between
the various possibilities in the near future.

One of the especially intriguing features of $R$-parity conserving
supersymmetric models is that they provide a natural candidate
for cold dark matter (CDM) in the universe\cite{haim,griest}.
The lightest neutralino $\tz_1$ in gravity-mediated SUSY breaking models
(SUGRA) is especially appealing as a DM candidate in that it can be
produced thermally in the early universe, and the 
calculable relic abundance turns out to be in the right neighborhood to
match the measurements of the density of CDM in the universe.

Many analyses of neutralino CDM have been performed\cite{wmapcon} 
within the context of the
paradigm minimal supergravity model\cite{msugra} (mSUGRA), which is
completely specified by the parameter set 
\be
m_0,\ m_{1/2},\ A_0,\ \tan\beta \ {\rm and} \ sign(\mu ).
\ee
The mSUGRA model assumes that the minimal supersymmetric 
Standard Model (MSSM) is
valid between the mass scales $Q=M_{GUT}$ and $Q=M_{\rm weak}$. A common
value $m_0$ ($m_{1/2}$) (($A_0$)) is assumed for all scalar mass
(gaugino mass) ((trilinear soft SUSY breaking)) parameters at
$Q=M_{GUT}$, and $\tan\beta$ is
the ratio of vacuum expectation values of the two Higgs fields that
give masses to the up and down type fermions.
The magnitude of the superpotential Higgs mass term $\mu$, but not its
sign, is fixed so as to reproduce the observed $Z$ boson mass.  The
values of couplings and other model parameters renormalized at the weak
scale can be computed via renormalization group (RG) evolution from
$Q=M_{GUT}$ to $Q=M_{\rm weak}$. From these weak scale parameters,
sparticle masses and mixings may
be computed, and the associated relic density of neutralinos as well as
scattering cross sections and decay rates can be determined.

In most of the allowed mSUGRA parameter space, 
the relic density $\Omega_{\tz_1}h^2$
turns out to be considerably larger than the WMAP value. Consistency
with WMAP thus implies that neutralinos should be able to annihilate
very efficiently in the early universe. In the mSUGRA model,
the annihilation rate is enhanced in 
just the following regions of parameter space, where the sparticle masses
and/or the neutralino composition assume special forms.
\begin{itemize}
\item The bulk region occurs at low values of $m_0$ and
$m_{1/2}$\cite{haim,bulk}.  In this region, neutralino annihilation is
enhanced by $t$-channel exchange of relatively light sleptons. The bulk
region, featured prominently in many early analyses of the relic
density, has been squeezed from below by the LEP2 bound on the chargino
mass $m_{\tw_1}>103.5$ GeV and the measured value of the branching
fraction $B(b\to s\gamma)$, and from above by the tight bound from WMAP.
\item The stau co-annihilation region  occurs at low $m_0$ for
almost any $m_{1/2}$ value where $m_{\ttau_1}\simeq m_{\tz_1}$. The
staus, being charged, can annihilate rapidly so that
$\ttau_1\tz_1$  co-annihilation processes
that maintain $\tz_1$ in thermal equilibrium with $\ttau_1$, serve to reduce
the relic density of neutralinos \cite{stau}.
\item The hyperbolic branch/focus point (HB/FP) region at large $m_0\sim$ 
several TeV, where $|\mu|$ becomes small, and neutralinos efficiently 
annihilate via their higgsino components\cite{hb_fp}.
This is the case of mixed higgsino dark matter (MHDM).
\item The $A$-annihilation funnel occurs at large $\tan\beta$ values
when $2m_{\tz_1}\sim m_A$ (or $m_H$)  and neutralinos can efficiently annihilate
through the relatively broad $A$ and $H$ Higgs resonances\cite{Afunnel}.
\end{itemize}
In addition, a less prominent light Higgs $h$ annihilation corridor occurs at
low $m_{1/2}$\cite{drees_h} 
and a top squark co-annihilation region occurs at 
particular $A_0$ values when $m_{\tst_1}\simeq m_{\tz_1}$\cite{stop}.

Many analyses have also been performed for gravity-mediated SUSY
breaking models with non-universal soft SUSY breaking (SSB) terms.
Non-universality of SSB scalar masses can, 1.~pull one or more scalar
masses to low values so that ``bulk'' annihilation via $t$-channel
exchange of light scalars can occur\cite{nmh,nuhm}, 2.~bring in
new near degeneracies of various sparticles with the $\tz_1$ so that new
co-annihilation regions open up\cite{auto,nuhm,sp}, 3.  bring the value
of $m_A$ into accord with $2m_{\tz_1}$ so that Higgs resonance
annihilation can occur\cite{ellis,nuhm}, or 4.~pull the value
of $|\mu|$ down so that higgsino annihilation can
occur\cite{ellis,drees2,nuhm}.

If non-universal gaugino masses are allowed, then qualitatively new
possibilities arise that are not realized in the mSUGRA
model\cite{ibanez,gunion,dermisek,models}.  One case, that of mixed wino
dark matter (MWDM), has been addressed in a previous paper\cite{winodm}.
In this case, as the weak scale value of $SU(2)$ gaugino mass $M_2({\rm
weak})$ is lowered from its mSUGRA value, keeping the hypercharge
gaugino mass $M_1({\rm weak})$ fixed, the wino component of $\tz_1$
continuously increases until it becomes dominant when $M_2({\rm weak}) <
M_1({\rm weak})$ (assuming $|\mu|$ is large). The $\tz_1\tw_{1,2}W$
coupling becomes large when $\tz_1$ becomes wino-like, resulting in
enhanced $\tz_1\tz_1\to W^+W^-$ annihilations.  Moreover,
co-annihilations with the lightest chargino and with the
next-to-lightest neutralino help to further suppress the thermal relic
abundance of the lightest SUSY particles (LSPs). Indeed, if the wino
component of the neutralino is too large, this annihilation rate is very
big and the neutralino relic density falls well below the WMAP value.

A qualitatively different case arises in supersymmetric models if the
SSB gaugino masses $M_1$ and $M_2$ are of opposite sign.
In this case, as $|M_1|$ and $|M_2|$ approach one another, 
there is little bino-wino mixing, and the $\tz_1$ maintains a 
nearly pure bino-like or wino-like identity. The WMAP relic 
density can nonetheless be achieved for $M_1\simeq -M_2$ via
bino-wino co-annihilation (BWCA) of the bino-like lightest neutralino.
The resulting DM and collider phenomenology was investigated in
Ref. \cite{binodm}. The MWDM and BWCA DM scenarios were also
investigated recently in Ref. \cite{nima}, 
where these scenarios were collectively dubbed 
``the well-tempered neutralino''.

In this paper, we investigate a scenario where, as previously noted
by Belanger {\it et al.}\cite{belanger} and also by Nezri and
Mambrini\cite{mn}, a diminution of the GUT scale value of the $SU(3)$
gaugino mass $M_3$ relative to $M_1$ and $M_2$ leads to a sparticle
spectrum with lower gluino and squark masses (the latter are lowered
through RG effects due to a reduced $M_3$). We are motivated to consider
this because by adjusting $M_3$ to the right value(s) provides another
way to obtain MHDM. To understand this, we begin by noting that the RG
equation for the soft SUSY breaking Higgs squared mass $m_{H_u}^2$ reads
(in a standard notation\cite{wss}) 
\be \frac{dm_{H_u}^2}{dt}
=\frac{2}{16\pi^2}\left( -{3\over 5}g_1^2M_1^2-3g_2^2M_2^2+{3\over
10}g_1^2 S+3f_t^2X_t\right) , 
\ee 
where
$X_t=m_{Q_3}^2+m_{\tst_R}^2+m_{H_u}^2+A_t^2$ and $S=m_{H_u}^2-m_{H_d}^2
+ Tr({\bf m}_Q^2- {\bf m}_L^2-2{\bf m}_U^2+{\bf m}_D^2 +{\bf m}_E^2)$.
Here, $f_t$ is the top quark Yukawa coupling and $t=\log Q^2$.  The
$f_t^2X_t$ term drives $m_{H_u}^2$ to negative values due to the large
top quark Yukawa coupling in the celebrated radiative electroweak
symmetry breaking (REWSB) mechanism\cite{rewsb}.  In the case where
$M_3<<M_1\sim M_2$ at the GUT scale, 
the squark squared mass terms and $A_t^2$ (and hence $X_t$)
are suppressed at lower scales; as a consequence, $m_{H_u}^2$ is not
driven to such large negative values as in the universal gaugino mass
case. Thus, if $|M_3|$ is chosen small (but not so small 
that $m_{H_u}^2$ is no longer
driven negative), we still obtain REWSB, but with a smaller weak scale
value of $-m_{H_u}^2$.\footnote{Indeed, for given values of other
parameters, the constraint of REWSB imposes a lower bound on $|M_3|$.}
There is also a corresponding effect on the RG flow of $m_{H_d}^2$, but
this is typically smaller because $f_b \ll f_t$ except for very large
values of $\tan\beta$. 
The weak scale value of $\mu^2$ (at tree-level) is then obtained 
from the weak scale parameters of the Higgs sector via the EWSB relation,
\be
\mu^2=\frac{m_{H_d}^2-m_{H_u}^2\tan^2\beta}{(\tan^2\beta -1)}-{M_Z^2\over 2} .
\ee
We see that if $|m_{H_u}^2| \gg M_Z^2$ and moderate to large 
values of $\tan\beta$, $\mu^2 \sim
-m_{H_u}^2$.  Thus the smaller $|m_{H_u}^2|$ value expected in the 
low $|M_3|$ case
results in a smaller $|\mu |$ parameter, and a correspondingly larger
higgsino component of the lightest neutralino $\tz_1$.

\FIGURE[!t]{
\epsfig{file=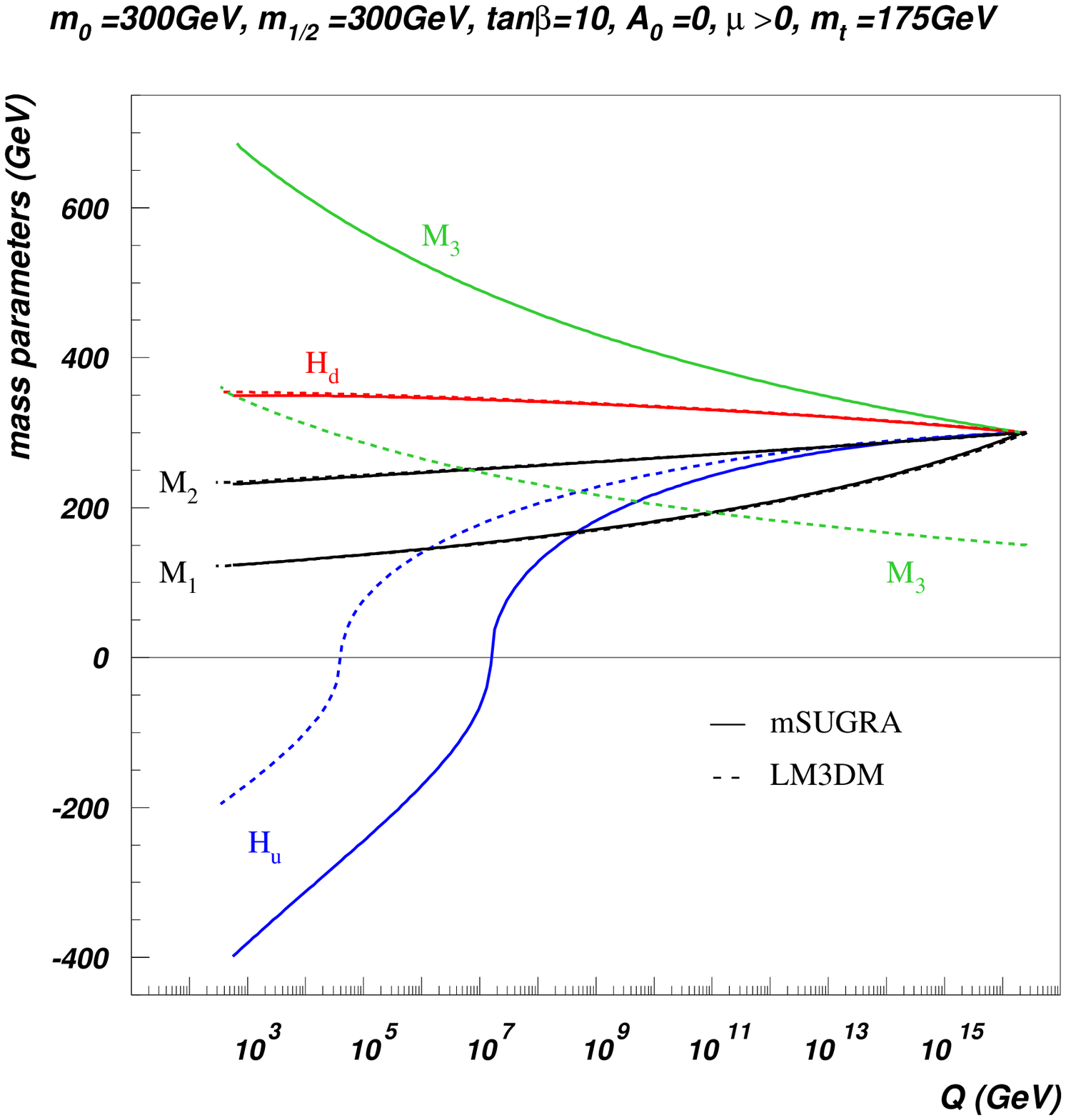,width=9cm} 
\caption{\label{fig:higgs_evol} Evolution of the soft SUSY breaking
Higgs mass parameters ${\rm sign}(m_{H_u}^2) \sqrt{|m_{H_u}^2}|$ 
and ${\rm sign} (m_{H_d}^2) \sqrt{|m_{H_d}^2}|$ as a function of scale
$Q$ in the mSUGRA model (solid) for $m_0=300$ GeV, $m_{1/2}=300$ GeV,
$A_0=0$, $\tan\beta =10$, $\mu >0$ and $m_t=175$ GeV. The same running
mass parameters are shown for LM3DM for the same parameters as in the
mSUGRA case except taking $M_3=0.5 m_{1/2}$ at $M_{GUT}$ (dashes). Also
shown is the corresponding evolution of gaugino mass parameters $M_1,
M_2$ and $M_3$, for the mSUGRA case (solid) and the $M_3 = 0.5m_{1/2}$
case (dashes).  }}
This situation is illustrated in Fig. \ref{fig:higgs_evol}, where we
plot the evolution of $m_{H_d}^2$, $m_{H_u}^2$, and the gaugino
mass parameters versus the
renormalization scale $Q$ from $Q=M_{GUT}$ to $Q=M_{\rm weak}$ for the
mSUGRA model with $m_0=m_{1/2}=300$ GeV, $A_0=0$, $\tan\beta =10$, $\mu
>0$ and $m_t=175$ GeV (solid curves), and for the case of $M_3=0.5
m_{1/2}$ (dashed curves).
The electroweak gaugino mass parameters evolve identically at the one
loop level, and the tiny difference seen is a two loop effect. The
gluino mass parameter, on the other hand, starts off at a smaller value
and evolves to a correspondingly smaller value at the weak scale.
Turning to the mass parameters in the Higgs sector, we see that, as
expected, $m_{H_u}^2$ runs to a less negative value in the case of the low
$M_3$ model than in the case of the mSUGRA model with universality of
the GUT scale gaugino mass parameters. The evolution of $m_{H_d}^2$ is
very similar in the two cases because the bottom Yukawa coupling is very
small. 

This reduction of $\mu$ is illustrated as well in Fig. \ref{fig:muvsm3},
where we show the behaviour of $\mu$ as a function of the ratio
$|M_3|/m_{1/2}$, for several values of $m_0$, again with $A_0=0$,
$\tan\beta =10$, $\mu >0$ and $m_t=175$ GeV. The curves end where REWSB
is no longer viable because $m_{H_u}^2$ does not evolve to negative
values. We see that for relatively low values of $m_0$ (smaller than a
few times $m_{1/2}$), low values of $\mu$ are achieved for
$|M_3|/m_{1/2}<1$, while for very large values of $m_0$ (that, for the
chosen value of $m_{1/2}$ may have
been forbidden for the mSUGRA case), REWSB with low values of $|\mu|$
becomes possible but only for $|M_3|/m_{1/2}>1$. 
These low $|\mu|$ regions are just generalizations of the well-known
HB/FP regions of the mSUGRA model.  
The location
of the ``generalized'' HP/FB region in
the $m_0-m_{1/2}$ plane of the extended model depends on the value of
$r_3 \equiv M_3/m_{1/2}$: it lies to the left (right) of the
corresponding region in the mSUGRA model if $|r_3| < 1$ ($|r_3|>1$). 
In the study presented here we focus on the first possibility, as
it leads to lighter coloured sparticles that may well be accessible at
the CERN Large Hadron Collider (LHC) scheduled to commence operations
next year. At the same time, the large higgsino component expected in
this low $M_3$ dark matter (LM3DM) scenario should lead to larger
detection rates relative to mSUGRA in direct and indirect searches for
neutralino dark matter.
%

%
\FIGURE[!t]{
\epsfig{file=nugm_mu.eps,width=9cm} 
\caption{\label{fig:muvsm3}
The values of $\mu$ dictated by REWSB as a function of $|M_3|/m_{1/2}$, for
$A_0=0$, $\tan\beta =10$, $\mu >0$
and $m_t=175$ GeV, at various values of $m_0$}}

Many previous studies have examined the neutralino relic density in
models with gaugino mass non-universality, along with prospects for
direct and indirect detection of DM neutralinos. 
Griest and Roszkowski\cite{gr}
first pointed out that a wide range of relic density values could be
obtained by abandoning gaugino mass universality by allowing departures
from $M_1/M_2 \simeq 0.5$.  Anomaly-mediated SUSY breaking
models, where the gaugino masses are proportional to the
$\beta$-functions of the corresponding low energy gauge groups have
$M_1:M_2:M_3\sim 3:1:-10$.  As a result, the $\tz_1$ is almost a pure
wino which annihilates very efficiently, resulting in too low a thermal
relic neutralino density: to account for the observed dark matter
density, Moroi and Randall\cite{moroi} invoked the decay of heavy moduli
to wino-like neutralinos in the early history of the universe.  The dark
matter relic density and detection rates in models with non-minimal
$SU(5)$ gauge kinetic function, and also in O-II string models were
studied by Corsetti and Nath\cite{cor_nath}.  Birkedal-Hanson and Nelson
showed that a GUT scale ratio $M_1/M_2\sim 1.5$ would bring the relic
density into accord with the measured CDM density via MWDM, and also
presented direct detection rates\cite{birkedal}.  Bertin, Nezri and
Orloff studied the variation of relic density and the enhancements in
direct and indirect DM detection rates as gaugino mass ratios are
varied\cite{nezri}. Bottino {\it et al.} performed scans over
independent weak scale parameters to show variation in indirect DM
detection rates, and noted that neutralinos as low as 6 GeV are
allowed\cite{bottino}. Mambrini and Mu\~noz, and also Cerdeno and
Mu\~noz, examined direct and indirect detection rates for models with
non-universal scalar and gaugino masses\cite{munoz}.  Auto {\it et
al.}\cite{auto} proposed non-universal gaugino masses to reconcile the
predicted relic density in models with Yukawa coupling unification with
the WMAP result.  Masiero, Profumo and Ullio exhibit the relic density
and direct and indirect detection rates in split supersymmetry where
$M_1$, $M_2$ and $\mu$ are taken as independent weak scale parameters
with ultra-heavy squarks and sleptons\cite{mpu}.  Finally, as mentioned
above, the variation of the relic density due to the change of $M_3$ --
the subject of this paper -- was first studied by Belanger {\it et al.}
who showed that large swaths of the $m_0-m_{1/2}$ plane are consistent
with the WMAP value when the $SU(3)$ gaugino mass $M_3$ becomes small
\cite{belanger}; this topic was subsequently also studied 
by Mambrini and Nezri\cite{mn}.

It has been shown that the
various non-universal scenarios each lead to distinctive phenomenologies,
and can be distinguished from mSUGRA and from one another via their
implications for accelerator experiments, and simultaneously, for direct
and indirect searches for DM. 
The purpose of this paper is to study WMAP viable SUSY models
with a non-universal GUT scale gaugino mass hierarchy $|M_3|\ll
M_1\simeq M_2$ -- these models have received relatively little attention
in the literature -- and to 
explore their phenomenology. 
In regions of parameter space that yield the observed relic density of
MHDM, we examine prospects for its direct and indirect detection, and
also outline the impact on prospects for direct detection of sparticles
at the Fermilab Tevatron, the CERN LHC and at the future international
linear $e^+e^-$ collider (ILC).  For expediency, we adopt an mSUGRA-like
model with universal GUT scale SSB parameters, but with the $SU(3)$
gaugino mass as one additional parameter; {\it i.e.} we take
$M_1=M_2\equiv m_{1/2}> 0$ at $Q=M_{GUT}$, while allowing $M_3(M_{GUT})$
to remain as a free parameter with either sign. By dialing $|M_3|$ to
low enough values (for the range of $m_0$ that we consider) any point in
the remainder of the parameter space can be WMAP allowed. The parameter
space naturally divides into regions with bino dark matter (BDM),
or with MHDM.
Once the WMAP constraint is fulfilled, then in the MHDM case one finds
generally enhanced rates for direct and indirect DM detection. As far as
colliders go, a mass spectrum with $m_{\tq}\simeq m_{\tell}$ is
predicted in the scalar sector. In the gaugino sector, a much reduced
mass gap of $m_{\tg}-m_{\tw_1}$ is expected as compared to mSUGRA.  This
means in part that lighter gluinos can be allowed despite the
constraints from LEP2, and that the Fermilab Tevatron may explore a
substantial portion of the LM3DM parameter space via gluino pair
production. We find that in the portion of the parameter space where
$m_{\tg}/M_2$ is most suppressed, $m_0$ is necessarily large, and
the radiative decays $\tg\to g\tz_i$ constitute the dominant 
decay modes of the gluino. In this case, gluino pair production may lead
to dijet $+\eslt$ events at hadron colliders.  At the CERN LHC, an
enhanced reach is found in $m_0\ vs.\ m_{1/2}$ parameter space relative
to the mSUGRA model due to the reduced squark and gluino masses.  At a
linear $e^+e^-$ collider, a much lighter spectrum of squarks and gluinos
is expected. In the case of MHDM, the low $\mu$ parameter implies that
the entire spectrum of charginos and neutralinos is rather light, and
may be accessible to ILC searches for new sparticles.

The remainder of this paper is organized as follows.
In Sec. \ref{sec:pspace}, we outline the parameter space of the 
LM3DM scenario, and show that any point in parameter space may be WMAP allowed
if a suitably low value of $|M_3|$ is adopted. We also illustrate
characteristic features of the sparticle mass spectrum expected in this
scenario. In Sec. \ref{sec:detect}, 
we discuss expectations for direct and indirect 
detection of neutralino DM in the LM3DM scenario, and show that
generally enhanced detection rates are expected if MHDM occurs.
In Sec. \ref{sec:col}, we give an overview of some of the main features
of the LM3DM scenario which give rise to distinct signatures at
the Fermilab Tevatron, the CERN LHC and a $\sqrt{s}= 0.5-1$ TeV ILC.
In Sec. \ref{sec:conclude}, we present a summary of our results 
and some conclusions.

{\it Note added:} As this paper was being finalized, the WMAP
collaboration released their three year data along with implications for
cosmology\cite{wmap3}. Their new central value of $\Omega_{\rm CDM} h^2$
(and of course, the quoted error) is slightly lower than in
(\ref{eq:wmap}). However, the central value obtained by combining the
WMAP data with other data as in Table~6 of Ref. \cite{wmap3} is almost
unchanged from (\ref{eq:wmap}). In either case, our analysis is hardly
affected.

\section{Parameter space and mass spectrum in the LM3DM scenario}
\label{sec:pspace}

As discussed in the previous section, the LM3DM scenario is completely
specified
by the parameter set:
\be
m_0,\ m_{1/2},\ M_3,\ A_0,\ \tan\beta ,\ {\rm and}\ sign(\mu ),
\ee
(together with $m_t$ which we fix to be 175~GeV), where we assume
$M_1=M_2\equiv m_{1/2}\ge 0$ at $Q=M_{GUT}$, and where $M_3$ can assume
either sign.  The assumed equality of $M_1$ and $M_2$ can be relaxed
somewhat and our conclusions suffer little qualitative change so long as
$M_1\simeq M_2$.  To calculate the sparticle mass spectrum, we adopt
Isajet 7.73\cite{isajet}, which allows for the input of non-universal
scalar and gaugino masses in gravity mediated SUSY breaking models where
electroweak symmetry is broken radiatively. The relic density is
evaluated via the IsaReD program\cite{isared}, which is part of the
Isatools package.  IsaReD evaluates all $2\to 2$ tree level neutralino
annihilation and co-annihilation processes and implements relativistic
thermal averaging in the relic density calculation.

We begin our discussion with an illustration to show that any point in
mSUGRA model parameter space that is WMAP disallowed owing to too large
a relic density $\Omega_{\tz_1}h^2$, can become WMAP allowed by
adjusting $|M_3|$ until a suitably small $|\mu |$ value that yields 
a relic density in accord with WMAP is attained. Assuming that
$m_0/m_{1/2}$ is not too large, this is achieved by {\it lowering}
$|M_3|$ from its mSUGRA value. As an example, we plot in
Fig. \ref{fig:rd}{\it a} the neutralino relic density
$\Omega_{\tz_1}h^2\ vs.\ M_3$ for $m_0=m_{1/2}=300$ GeV, while $A_0=0$,
$\tan\beta =10$ and $\mu >0$.  The value $M_3=300$ GeV puts us in the
mSUGRA model, and here we see $\Omega_{\tz_1}h^2=1.1$, and so the
parameter space point would be excluded. As $M_3$ is lowered from its
mSUGRA value, gluino and squark masses also decrease, resulting in a
lower weak scale value of $|m_{H_u}^2|$ and hence $|\mu |$ as discussed in
Sec. \ref{sec:intro}. At $M_3=150$ GeV, the value of $|\mu |$ has
diminished sufficiently that the $\tz_1$ is no longer bino-like, but is
instead a mixed higgsino-bino state. This is illustrated in frame {\it
b}), where we plot $R_{\tH} =\sqrt{v_1^{(1)2}+v_2^{(1)2}}$ (in the
notation of Ref. \cite{wss}), which gives an indication of the higgsino
components of the $\tz_1$. As $M_3$ is decreased even further, the relic
density reaches a minimum around $M_3\sim 110$ GeV, and then increases
slightly before reaching the LEP2 limit where $m_{\tw_1}$ becomes less
than 103.5 GeV. This slight increase occurs because $m_{\tz_1}$ drops
below $M_Z$ and then $M_W$, so that $\tz_1\tz_1\to ZZ,\ W^+W^-$
processes, the major LSP annihilation modes in the early universe become
kinematically suppressed.  
Since there is no reason to favour the positive sign of $M_3$, 
we show the behavior of the relic density and $R_{\tH}$ for
negative $M_3$ as well, and note that these
are nearly symmetrical under 
$M_3 \to -M_3$.
\FIGURE[!t]{
\epsfig{file=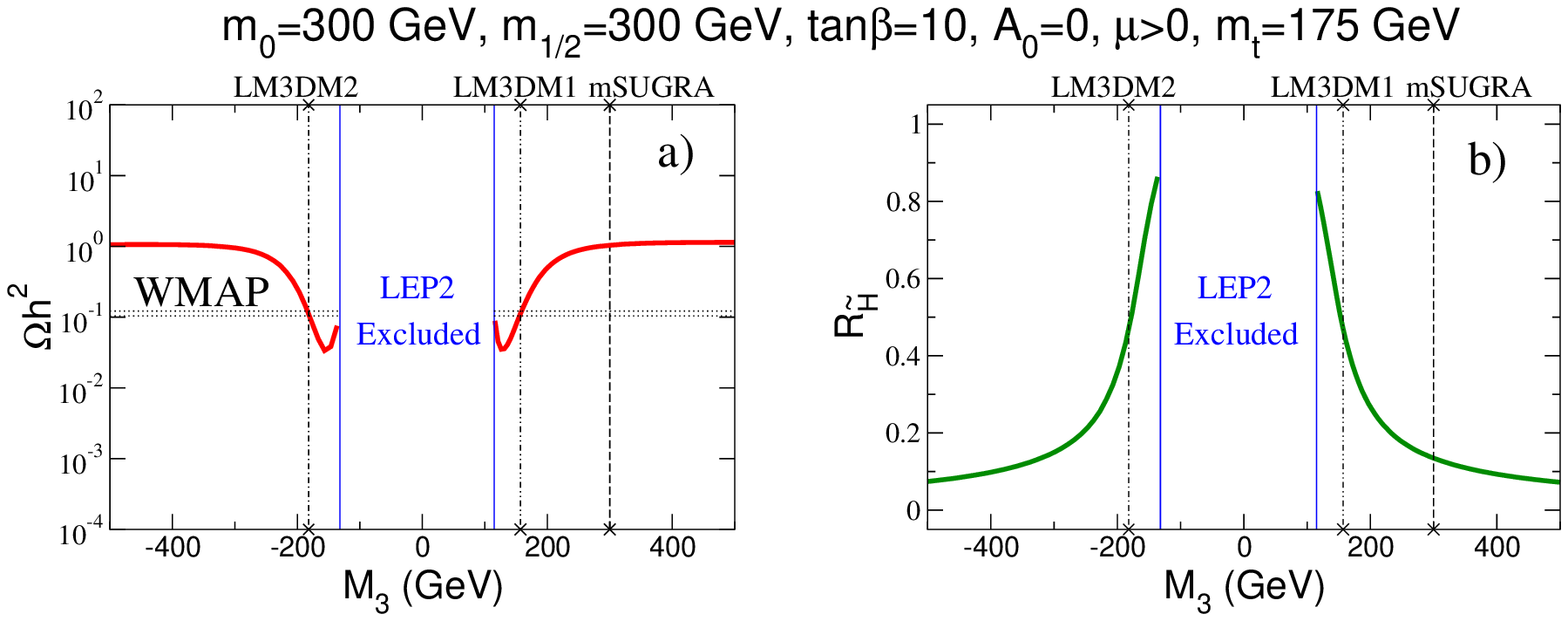,width=7cm,angle=-90} 
\caption{\label{fig:rd}
{\it a}) The neutralino relic density $\Omega_{CDM}h^2$, and 
{\it b}) higgsino component $R_{\tH}$ of the lightest neutralino as a
function of $M_3$ for
$m_0=300$ GeV, $m_{1/2}=300$ GeV, $A_0=0$, $\tan\beta =10$, $\mu >0$
and $m_t=175$ GeV.}}

The effect on the sparticle mass spectrum of lowering the magnitude of
the $SU(3)$ gaugino mass $M_3$ is shown in Fig.~\ref{fig:mass}, where we
plot the sparticle mass spectrum versus the ratio $r_3=M_3/m_{1/2}$ for
the same parameters as in Fig. \ref{fig:rd}. At $r_3 =1$, we see
the usual hierarchy of sparticle masses as obtained in the mSUGRA
model. As $M_3$ is lowered, the gluino masses reduces sharply from
$m_{\tg}= 727$ GeV in the mSUGRA case to $m_{\tg}\simeq 400$ GeV for
$r_3 =0.5$ where $\Omega_{\tz_1}h^2= 0.11$. The reduction of
$M_3$ reduces the renormalization suffered by the squark mass
parameters, and causes the squark masses to correspondingly drop from
the vicinity of 700 GeV in mSUGRA to $\sim 500$ GeV for
$r_3 =0.5$. On the other hand, slepton mass parameters, whose
renormalization does not depend on SUSY QCD effects at the one loop
level, are hardly affected by the change of $M_3$. Thus, in the LM3DM
scenario, the mass scale of squarks and sleptons is more nearly
equal, and less hierarchical, than the case of mSUGRA at low $m_0$.  In
fact, for the case shown in Fig. \ref{fig:mass}, the top squark $\tst_1$
has dropped to a lower mass than the various sleptons in the case of
$r_3 =0.5$. The value of the $\mu$ parameter is shown by the
black dotted curve, and we see that this drops sharply as $|M_3|$
decreases. The drop in $|\mu|$ increases the higgsino component of the
lighter charginos and neutralinos and, once they cross over to becoming
higgsino-like, their masses decrease with decreasing $M_3$ as well.
\FIGURE[!t]{
\epsfig{file=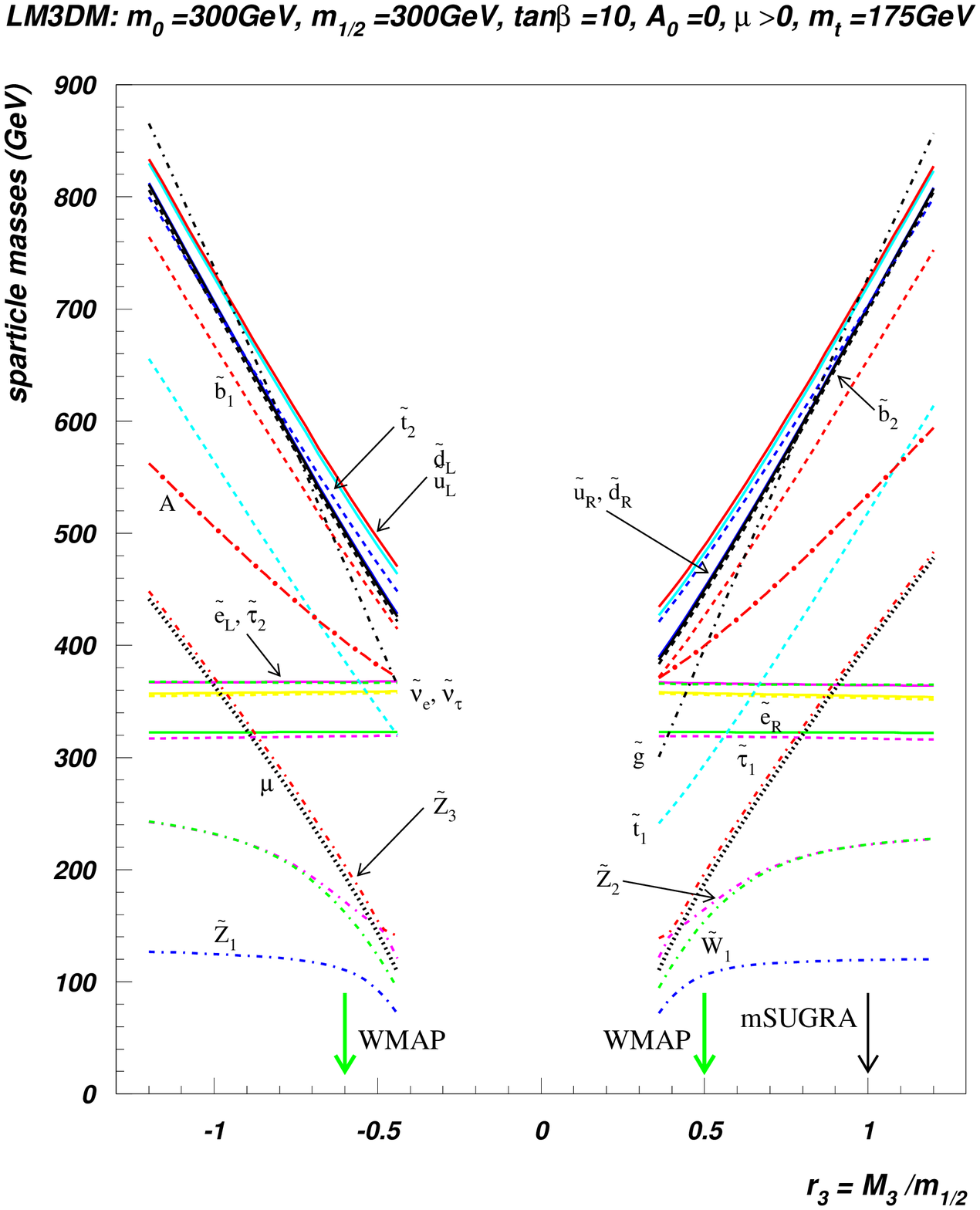,width=10cm} 
\caption{\label{fig:mass} A plot of various sparticle and Higgs boson
masses and the $\mu$ parameter {\it vs.} $M_3/m_{1/2}$ for $m_0=300$
GeV, $m_{1/2}=300$ GeV, $A_0=0$, $\tan\beta =10$ and $\mu >0$.}}

We also show in Fig. \ref{fig:mass} the sparticle masses for negative 
values of $r_3$. The slepton and first/second generation
squark masses are nearly symmetrical about $M_3=0$ because the leading
(one loop) contributions to their renormalization group evolution are
quadratic in the gaugino masses.
However, the top squark and various chargino and neutralino 
masses are not symmetric. For the stops, this occurs
because the RG evolution of the $A$ parameters 
($A_t$ in this case)
depends linearly on the gaugino masses. The asymmetric evolution of the
$A$-parameters also affects  $m_{H_u}^2$ and $m_{H_d}^2$ the same way;
as a result, $|\mu|$, and hence chargino and neutralino
masses are also not symmetric under $M_3 \to -M_3$. 
We see, in fact, when $M_3<0$, the 
WMAP measured value of $\Omega_{CDM}h^2$ is attained at a value
of $r_3 =-0.6$, in contrast to $r_3 =0.5$ for positive
values of $M_3$. 

Various sparticle masses are shown in Table \ref{tab:m3dm} corresponding
to the parameters shown in Fig. \ref{fig:mass}.  We show the spectrum
for the mSUGRA case, together with that for the LM3DM1 case (with
$r_3=0.5$) and the LM3DM2 case, with $r_3=-0.6$. In the several rows
(below the masses) we show the relic density, $BF(b\to s\gamma )$,
$\Delta a_\mu$, and the neutralino-proton scattering cross section
$\sigma (p\tz_1 )$ for these scenarios.  Finally, in the last row, we
show that component of the ``up-type'' neutral higgsino, {\it i.e.} that
couples to the $T_3=+1/2$ quark-squark system, in the neutralino LSP: we
will return to this, as well as the LM3DM scenario in the last column
of Table \ref{tab:m3dm} 
when we discuss Tevatron signals for the gluino in the LM3DM
framework.\footnote{We are aware that the value of $m_h$, especially in
the LM3DM1 and LM3DM2 scenarios, is well below the limit from searches
for the SM Higgs boson which should be applicable in these cases because
$m_A$ is large. Since the value of $m_h$ depends on our choice of
$\tan\beta$ as well of $A_0$ (whose precise values do not qualitatively
affect the features that we discuss here), we will continue to use these
scenarios as simple illustrations of the model.}

%
\begin{table}
\begin{tabular}{lcccc}
\hline
parameter & mSUGRA & LM3DM1 & LM3DM2 & LM3DM3 \\
\hline
$m_0$ & 300 & 300 & 300 & 1500 \\
$M_1$ & 300 & 300 & 300 & 250 \\
$M_2$ & 300 & 300 & 300 & 250 \\
$M_3$ & 300 & 150 & -180 & 50 \\
$\mu$ & 400.5 & 187.9 & 194.0 & 140.7 \\
$m_{\tg}$ & 727.3 &  396.7 & 472.6 & 182.7 \\
$m_{\tu_L}$ & 720.7 & 482.4 & 533.3 & 1492.2 \\
$m_{\tst_1}$ & 518.4 & 293.7 & 385.9 & 838.9 \\
$m_{\tb_1}$ & 654.6 & 426.7 & 482.1 & 1213.6 \\
$m_{\te_L}$ & 364.6 & 366.2 & 367.6 & 1506.2 \\
$m_{\te_R}$ & 322.3 & 322.6 & 322.7 & 1501.7 \\
$m_{\tw_2}$ & 425.2 & 283.9 & 292.3 & 249.7 \\
$m_{\tw_1}$ & 222.5 & 154.2 & 161.8 & 115.9 \\
$m_{\tz_4}$ & 426.0 & 286.2 & 294.7 & 254.8 \\
$m_{\tz_3}$ & 406.3 & 196.8 & 203.8 & 153.0 \\ 
$m_{\tz_2}$ & 222.3 & 164.9 & 171.4 & 133.1 \\ 
$m_{\tz_1}$ & 119.5 & 106.2 & 110.5 & 81.0 \\ 
$m_A$       & 533.5 & 400.0 & 403.6 & 1496.6 \\
$m_{H^+}$   & 542.9 & 410.4 & 414.0 & 1508.6 \\
$m_h$       & 110.7 & 106.1 & 103.9 & 110.2 \\ \hline
$\Omega_{\tz_1}h^2$& 1.1 & 0.11 & 0.12 & 0.13 \\
$BF(b\to s\gamma)$ & $3.1\times 10^{-4}$ & $1.6\times 10^{-4}$ &
$5.5\times 10^{-4}$ & $3.4\times 10^{-4}$ \\
$\Delta a_\mu    $ & $11.9 \times  10^{-10}$ & $16.3 \times  10^{-10}$ & 
$15.7\times 10^{-10}$ & $1.5\times 10^{-10}$ \\ 
$\sigma_{sc} (\tz_1p )$ & 
$1.8\times 10^{-9}\ {\rm pb}$ & $6.8\times 10^{-8}\ {\rm pb}$ & 
$6.7\times 10^{-8}\ {\rm pb}$ & $4.2\times 10^{-8}\ {\rm pb}$\\
$|v_1^{(1)}|$ & 0.05 & 0.26 & 0.26 & 0.37 \\
\hline
\end{tabular}
\caption{Masses and parameters in~GeV units
for mSUGRA and three LM3DM scenarios. In each case,
$A_0=0$, $\tan\beta =10$ and $m_t=175$ GeV.
}
\label{tab:m3dm}
\end{table}

In Fig. \ref{fig:r3}, we show contours of $r_3$ in the $m_0\ vs.\ m_{1/2}$
plane for $A_0=0$, $\tan\beta =10$ and $\mu >0$, where at each point 
in the plane $|M_3|$ has been reduced until the value of the relic density
is found to be $\Omega_{\tz_1}h^2=0.11$. 
Frame {\it a}) shows these contours for
negative $r_3$, while frame {\it b}) shows contours for $r_3>0$. 
The red shaded region on the left hand side is excluded 
because the $\ttau_1$ becomes the LSP (in contradiction to search limits for
stable charged or colored relics), while the blue region at low $m_{1/2}$
is excluded by the LEP2 chargino search limits, which require 
$m_{\tw_1}>103.5$ GeV. Unlike in the mSUGRA model, the LEP-excluded blue 
region is not flat because the chargino is gaugino-like for small values
of $m_0$ but higgsino-like as $m_0$ becomes large. 
Of course, in the stau co-annihilation region, which is tight against the
boundary of the red region, the relic density is already 
in accord with the WMAP value even in the mSUGRA case. 
In the lower-left bulk region, $|M_3|$
need only be reduced to values of $r_3\sim - 0.7$ (frame {\it a})) or
$r_3\sim 0.6$ (frame {\it b})). However, for large $m_0$ and low $m_{1/2}$,
values of $|r_3|$ as low as $\sim 0.3$ are needed to reach accord with WMAP.
A third case study in this region, labelled LM3DM3, is shown in the last
column of 
Table \ref{tab:m3dm}.\footnote{In this case, the decays $\tz_4\to
  q\bar{q}\tg$ and the decays $\tw_2^+ \to u\bar{d}\tg$, not included in
  Isajet, are kinematically accessible. We expect, however, that this
  will not significantly affect our analysis because these sparticles
  will dominantly decay via their two-body modes.}
In most of parameter space, values of $|r_3|\sim 0.4-0.6$ are sufficient
to match the measured relic density. 
\FIGURE[!t]{
\mbox{\epsfig{file=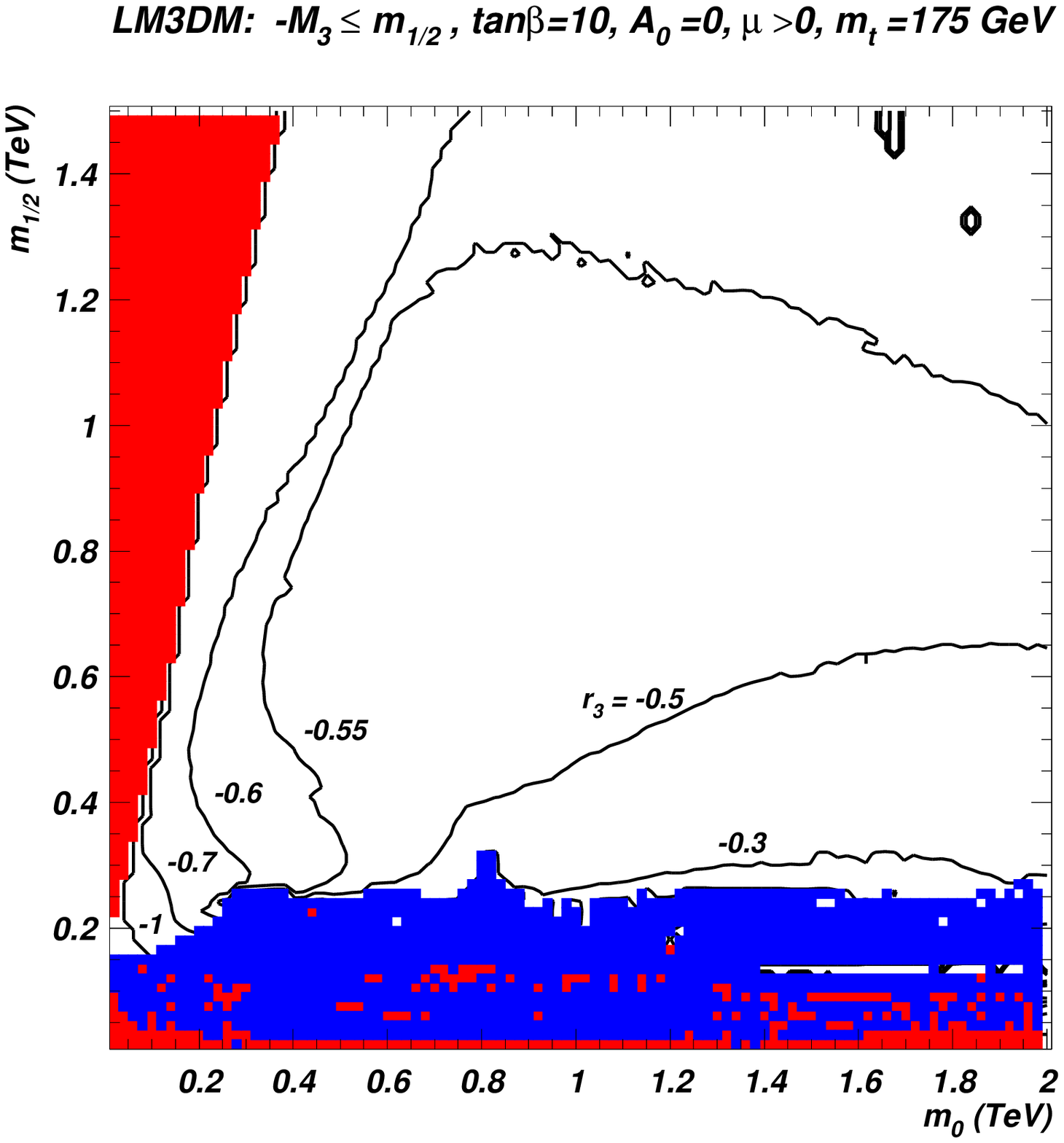,width=7cm}
\epsfig{file=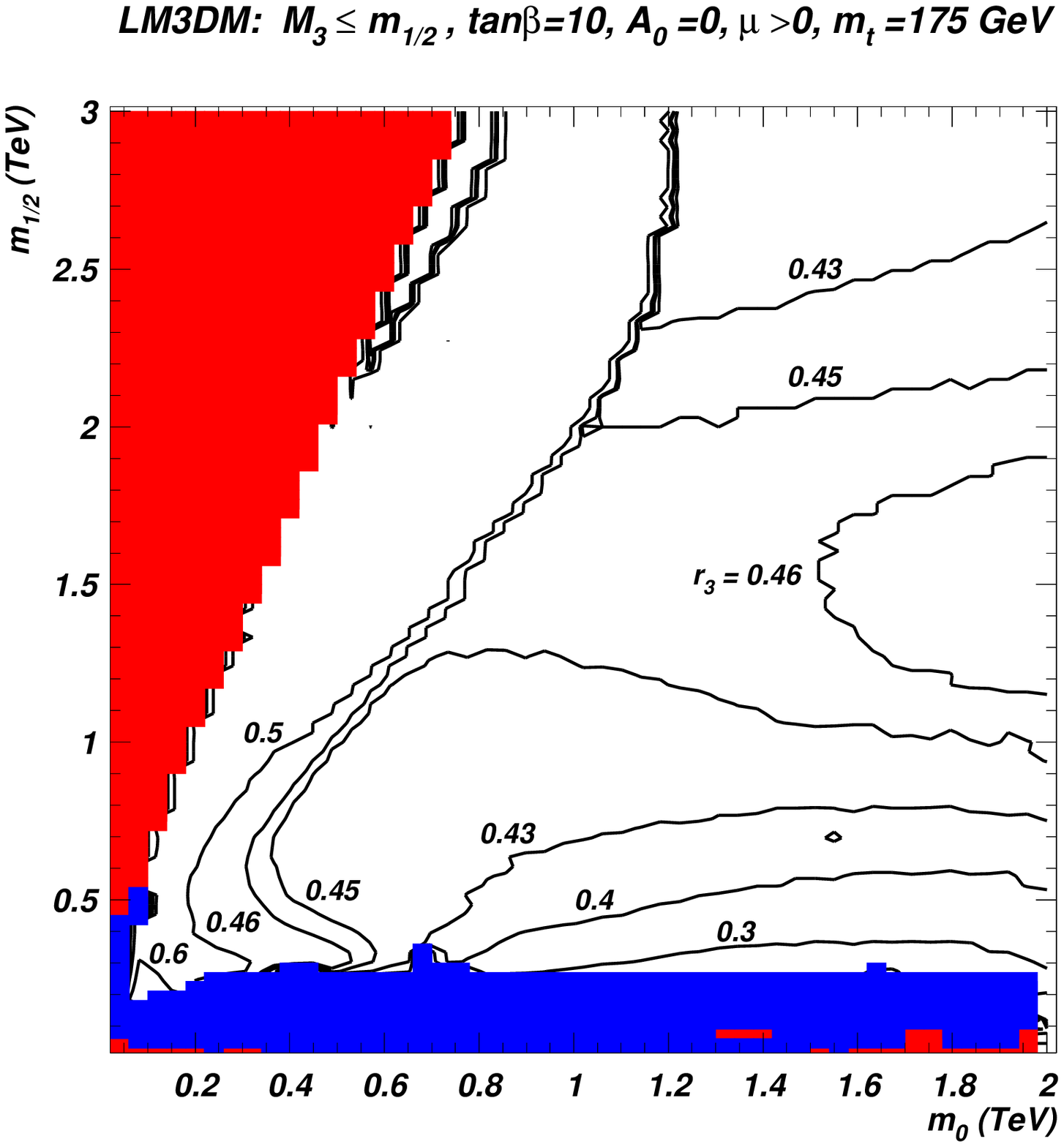,width=7cm}}
\caption{\label{fig:r3}
Contours of {\it a}) $r_3<0$ in the 
$m_0\ vs.\ m_{1/2}$ plane for $m_{1/2}$ up to 1.5 TeV, with
$\tan\beta =10$, $A_0=0$, $\mu >0$. Each point in the plane has
$r_3$ dialed to such a value that $\Omega_{\tz_1}h^2 =0.11$. The red
region on the left is excluded either because $\ttau_1$ becomes the LSP
or electroweak symmetry is not correctly broken, while the blue region
is excluded by the LEP lower limit $m_{\tw_1}> 103.5$~GeV. 
In frame {\it b}), we plot contours of $r_3>0$ for the same parameter choices
as in frame {\it a}), although we extend the range of $m_{1/2}$ to 3 TeV.
}}

The planes of Fig. \ref{fig:r3} naturally divide into two distinct
regions. The left-hand side of each plot with $|r_3|\agt 0.5-0.6$ is labelled
the bino DM region (BDM), since here the $\tz_1$
is bino-like, while the large $m_0$ side of the plane with 
$|r_3|\alt 0.5-0.6$ is labelled as 
MHDM, since here the $\tz_1$ is mixed higgsino-bino state. The
bino-wino-higgsino content of $\tz_1$ for an $m_{1/2}=1005$ GeV slice out
of Fig. \ref{fig:r3} is shown in Fig. \ref{fig:R1005}.
\FIGURE[!t]{
\epsfig{file=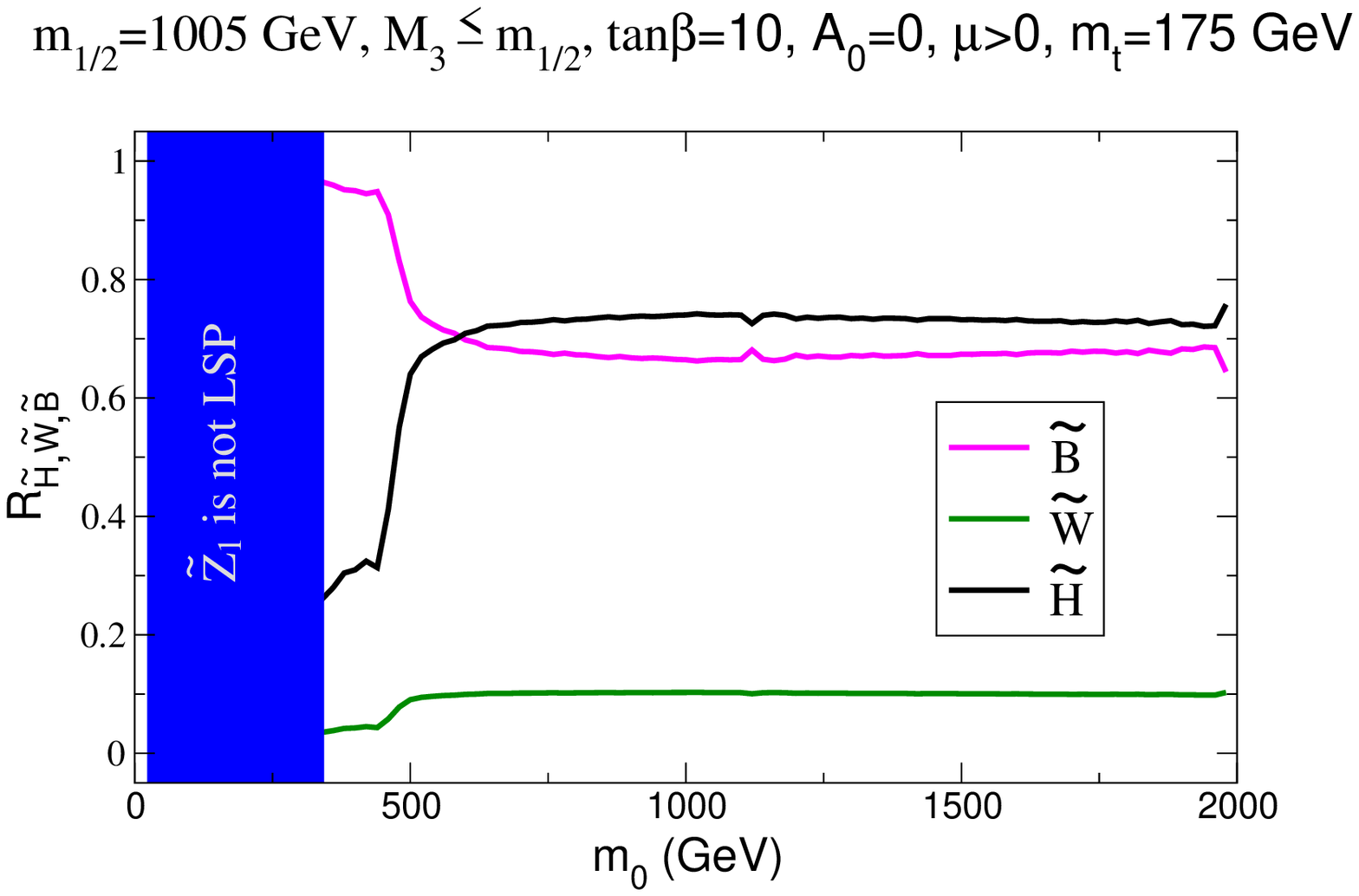,width=6cm,angle=-90}
\caption{\label{fig:R1005}
The bino, wino and higgsino content of the $\tz_1$ versus $m_0$ for
a $m_{1/2}=1005 $ GeV slice out of the plane in Fig. \ref{fig:r3}{\it b}),
showing that $\tz_1$ is BDM for small $m_0$, and MHDM for large $m_0$.
}}

In the BDM regions of Fig. \ref{fig:r3}, the values of $M_3$ and $m_0$
are low enough that $m_A$ approaches $2m_{\tz_1}$. In this case, the
thermal distribution of neutralinos convoluted with the $\tz_1\tz_1\to
b\bar{b}$ cross section allows for an enhanced annihilation rate via the
$s$-channel $A,\ H$-pole. On the very low $m_0$ edge of parameter space,
$\ttau_1$ co-annihilation also contributes to a reduction in the
neutralino relic density, and a wide range of $r_3$ is possible. As we move to larger $m_0$ values in
Fig. \ref{fig:r3}, the value of $m_A$ becomes much larger than
$2m_{\tz_1}$, and $A$-funnel annihilation is no longer efficient enough
to reduce the relic density.  Various sparticle and Higgs masses are
shown as a function of $m_0$ in Fig. \ref{fig:m1005}, for the same
parameters as in Fig. \ref{fig:R1005}.  In this case, for large $m_0$,
$r_3$ must be reduced to lower values of $r_3<0.5$, and the $\tz_1$
becomes MHDM. Then $\tz_1\tz_1\to WW$, $ZZ$, and also $\tw_1\tz_1$ and
(to a smaller extent) $\tz_1\tz_2$ co-annihilation act to suppress the
relic density. This is shown explicitly in Fig. \ref{fig:sigv1005}, where we
plot the thermally averaged neutralino annihilation cross sections
integrated over temperature from freeze-out to the present time, versus
$m_0$ for the same parameter choices as in Fig. \ref{fig:R1005}.
We see that the
co-annihilation processes become significant only for $m_0 \agt 400$~GeV
where $r_3$, and correspondingly also $|\mu|$, have become sufficiently
small.
\FIGURE[!t]{
\epsfig{file=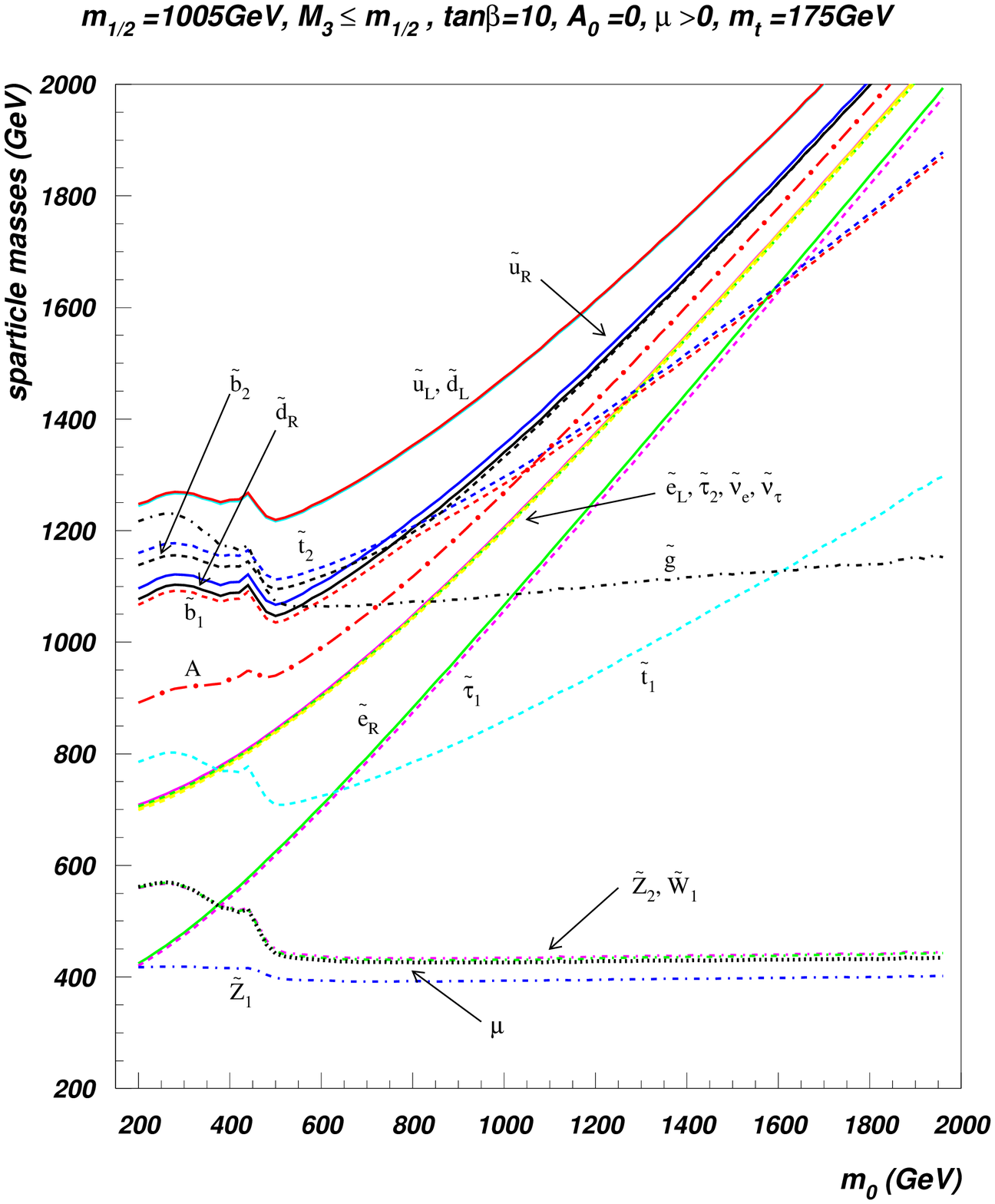,width=9cm}
\caption{\label{fig:m1005}
Sparticle masses vs. $m_0$ for
an $m_{1/2}=1005 $ GeV slice out of the plane in Fig. \ref{fig:r3}{\it b}).
}}
\FIGURE[!t]{
\epsfig{file=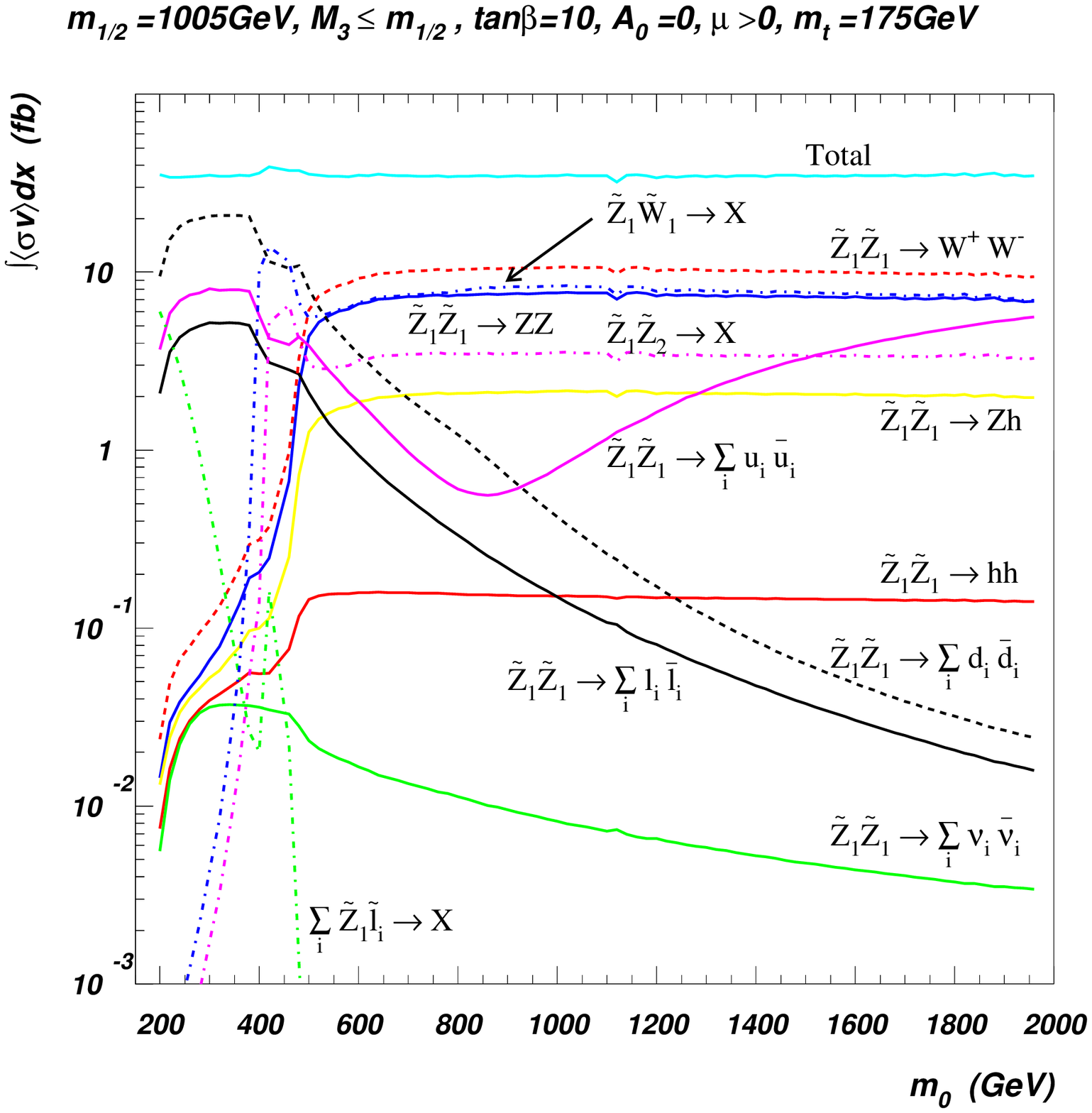,width=9cm}
\caption{\label{fig:sigv1005}
The thermally averaged neutralino annihilation cross sections
times relative velocity integrated from $x=0$ to $x_F$ versus $m_0$ for
a $m_{1/2}=1005 $ GeV slice out of the plane in Fig. \ref{fig:r3}{\it b}).
}}

Finally, we mention that in the upper-left of frame of
Fig. \ref{fig:r3}{\it b}) (which has been extended to $m_{1/2}=3$ TeV to
facilitate the discussion of the LHC reach in Sec. \ref{ssec:lhc}), 
the $r_3$ value drops below 0.5 near the contours at $m_0 \sim 0.8$~TeV
and $m_{1/2}\sim 3$~TeV. In this part
of the plane, a small additional reduction in $M_3$ is needed to match
$\Omega_{CDM}h^2=0.11$ because the enhancement to
$\tz_1\tz_1$ annihilation via amplitudes with  $s$-channel
$H,\ A$  exchanges no longer obtains
since $2m_{\tz_1}>m_H,\ m_A$. Thus, in this upper-left
region, the $\tz_1$ is again MHDM.

\section{Direct and indirect detection of neutralino CDM}
\label{sec:detect}

In this section, we explore the prospects for direct and indirect
detection of neutralino dark matter within the LM3DM framework
\cite{eigen}.  We adopt the DarkSUSY code\cite{darksusy}, interfaced to
Isajet, for the computation of the various indirect detection 
rates, and resort to the
Adiabatically Contracted N03 Halo model\cite{n03} for the dark matter
distribution in the Milky Way, which tends to give higher detection
rates, especially for gamma ray and anti-particle detection than other
halo profiles. In this respect, our projections may be regarded as
optimistic.\footnote{For a comparison of the implications of different
halo model choices for indirect DM detection rates, see {\it e.g.}
Refs. \cite{bo,bbko,antimatter,nuhm}.}  We evaluate the following
neutralino DM detection rates:
\begin{itemize}
\item Direct neutralino detection via underground cryogenic
detectors\cite{direct}.  Here, we compute the spin independent
neutralino-proton scattering cross section, and compare it to expected
sensitivities\cite{bbbo} for Stage 2 detectors (CDMS2\cite{cdms2},
Edelweiss2\cite{edelweiss}, CRESST2\cite{cresst}, ZEPLIN2\cite{zeplin})
and for Stage 3, SuperCDMS, along with ton-size detectors (XENON\cite{xenon},
GERDA\cite{gerda}, ZEPLIN4\cite{zeplin4} and WARP\cite{warp}).  We adopt
the projected (mass-dependent) sensitivities of CDMS2 and 1-ton XENON
detectors as the experimental benchmark for direct DM detection at
stage~2 and stage~3 detectors.


\item Indirect detection of neutralinos via neutralino annihilation to
neutrinos in the core of the Sun\cite{neut_tel}. 
Here, we present rates for detection of $\nu_\mu \to \mu$ conversions
at Antares\cite{antares} or IceCube\cite{icecube}. 
The reference experimental sensitivity we use is that of IceCube, 
with a muon energy threshold of 50 GeV, corresponding to a flux 
of about 10 muons per ${\rm km}^2$ per year. 
\item Indirect detection of neutralinos via neutralino annihilation in the
galactic center leading to gamma rays\cite{gammas}, 
as searched for by EGRET\cite{egret}, and 
in the future by GLAST\cite{glast}. 
We evaluate the integrated continuum $\gamma$ ray flux above a 
$E_\gamma=1$ GeV threshold, and assume a GLAST sensitivity 
of 1.0$\times10^{-10}\ {\rm cm}^{-2}{\rm s}^{-1}$.
\item Indirect detection of neutralinos via neutralino annihilations in
the galactic halo leading to cosmic antiparticles, including
positrons\cite{positron} (HEAT\cite{heat}, Pamela\cite{pamela} and
AMS-02\cite{ams}), antiprotons\cite{pbar} (BESS\cite{bess}, Pamela,
AMS-02) and anti-deuterons ($\bar{D}$) (BESS\cite{bessdbar}, AMS-02,
GAPS\cite{gaps}).  For positrons and antiprotons we evaluate the
averaged differential antiparticle flux in a projected energy bin
centered at a kinetic energy of 20 GeV, where we expect an optimal
statistics and signal-to-background ratio at space-borne antiparticle
detectors\cite{antimatter,statistical}. We take the experimental
sensitivity to be that of the Pamela experiment after three years of
data-taking as our benchmark.  Finally, the average differential
antideuteron flux has been computed in the $0.1<T_{\bar D}<0.25$ GeV
range, where $T_{\bar D}$ stands for the antideuteron kinetic energy per
nucleon, and compared to the estimated GAPS sensitivity for an
ultra-long duration balloon-borne experiment \cite{gaps} (see
Ref.~\cite{baerprofumo} for an updated discussion of the role of
antideuteron searches in DM indirect detection).
\end{itemize}

In Fig.~\ref{dmrates1}, we show various direct and indirect DM detection
rates for $m_0=m_{1/2}=300$ GeV, with $A_0=0$, $\tan\beta =10$ and $\mu
>0$, while $M_3$ is allowed to vary.  The $M_3$ value corresponding to
the mSUGRA model is denoted by a dashed vertical line, while the LM3DM
scenarios for $r_3<0$ and $r_3>0$ with $\Omega_{\tz_1}h^2=0.11$ are
denoted by dot-dashed and dot-dot-dashed vertical lines, respectively.
The dotted lines correspond to the sensitivity level of each of these
experiments; {\it i.e.}, the signal is observable only when the model
prediction is higher than the corresponding dotted line. While the
minimum sensitivity for the direct detection rates in frames {\it b}) --
{\it f}) refers to the minimum magnitude of the signal that is
detectable (and hence independent of the LSP mass), the smallest
detectable cross section shown by the dotted curves in frame {\it a})
depends on the value of $m_{\tz_1}$.

In frame {\it a}), we plot the spin-independent neutralino-proton
scattering cross section.  We see that as $M_3$ is decreased from its
mSUGRA value, the neutralino-proton scattering cross section rises
almost two orders of magnitude to a value above $3\times 10^{-8}$ pb,
which should be detectable by CDMS2, and certainly at stage 3 detectors.
A similarly large rate is attained for $r_3<0$, as shown in the
left-hand side of frame {\it a}). This frame merely reflects the
well-known result that MHDM has rather large neutralino-proton
scattering rates, as is typified by the HB/FP region of the mSUGRA
model. The value of $\sigma_{SI}(\tz_1 p)$ is further enhanced by the 
lowered squark masses of the LM3DM senario.
\FIGURE[!t]{
\mbox{\hspace{-1cm}
\epsfig{file=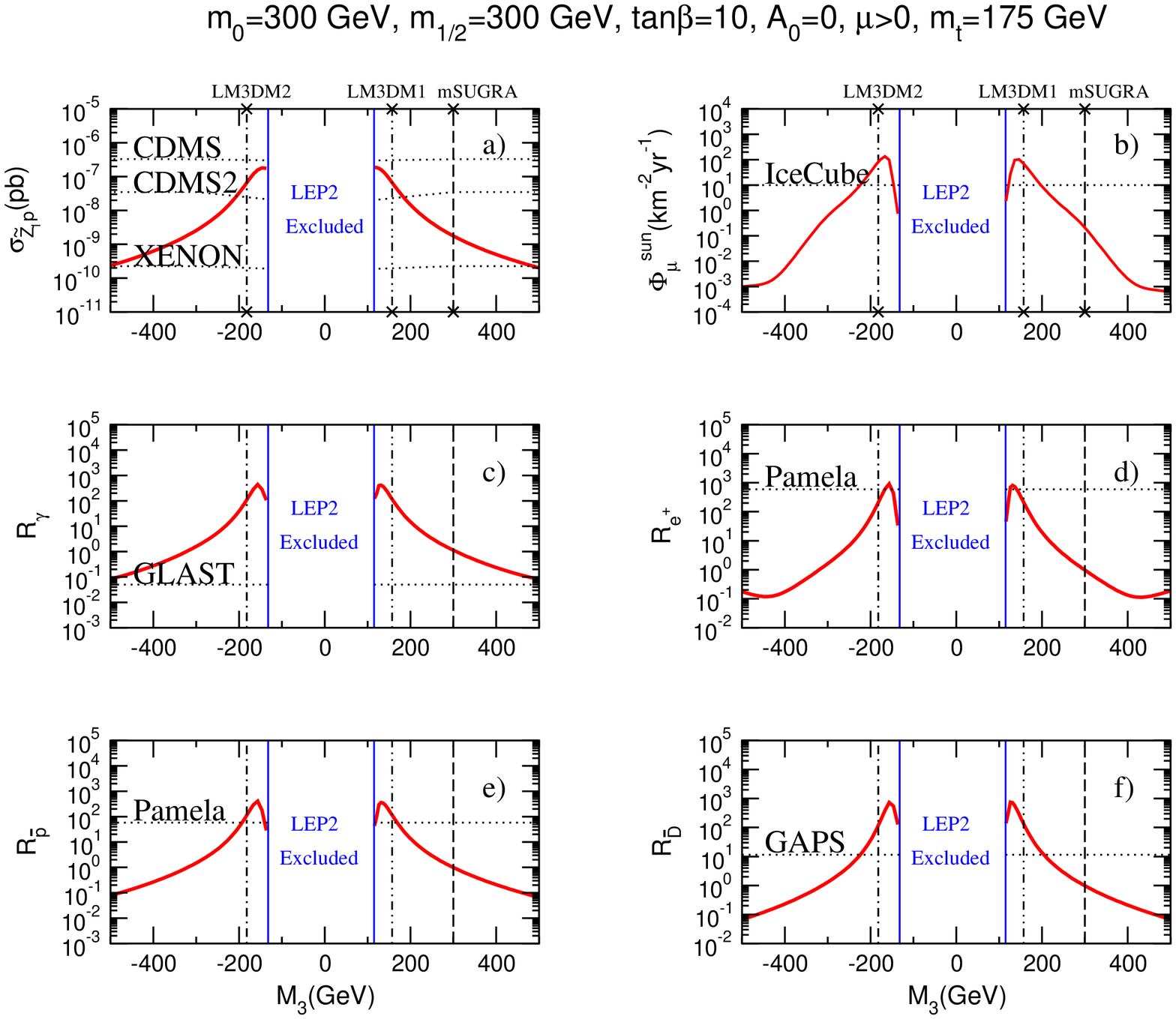,width=12cm,angle=-90} }
\caption{\label{dmrates1}
Rates for direct and indirect detection of neutralino dark matter
vs. $M_3$ for $m_0=m_{1/2}=300$ GeV, with
$\tan\beta =10$, $A_0=0$, $\mu >0$. 
Frames {\it c}) -{\it f}) show the ratio of indirect detection rates 
compared to the mSUGRA model. 
In this figure, we adopt the adiabatically contracted 
N03 distribution for halo dark matter for our projections of the reach
of the various experiments.}}

In frame {\it b}), we show the flux of muons from neutralino pair
annihilations in the core of the Sun.  The expected muon flux is below
the reach of IceCube in the mSUGRA framework, but increases over two
orders of magnitude into the observable range for IceCube as a result of
the increased
higgsino content of the LSP and the reduced squark mass when the relic
density is in agreement with the WMAP measurement as in the LM3DM model.  

In frames {\it c}), {\it d}), {\it e}) and {\it f}) we show the flux of
photons, positrons, antiprotons and antideuterons, respectively.  The
results here are plotted as ratios of fluxes normalized to the
corresponding mSUGRA
point, in order to give results that are approximately halo-model
independent.  Also  shown by the horizontal
lines are the expected experimental
reaches, as obtained by using the Adiabatically Contracted N03 Halo
model\cite{n03}.   The rates for indirect detection via observation of
halo annihilation remnants are typically low for bino-like DM as
in the mSUGRA model. However, when $|r_3|$ is reduced until
the measured CDM relic abundance is achieved, these halo annihilation
signals all jump by factors of 100-200, and are much more likely to be
observed by various gamma ray and antimatter detection experiments. We
should, however, keep in mind that this conclusion is sensitive to our
assumption of the DM halo profile.

The detectability of the same signals, but this time for the slice of
the $m_{1/2}=1005$~GeV slice of LM3DM parameter space we considered in
Fig.~\ref{fig:m1005} is illustrated in Fig.~\ref{fig:dmdet1005}. The
cross sections or the expected fluxes are absolutely normalized, rather
than to any particular mSUGRA model. The most striking feature of this
figure is the rather sudden increase (around $m_0\sim 500$~GeV) of the
signal as $m_0$ increases from low values where the LSP is bino-like to
high values where the required low value of $r_3$ leads to a significant
higgsino component in the LSP. 

\FIGURE[!t]{
\mbox{\hspace{-1cm}
\epsfig{file=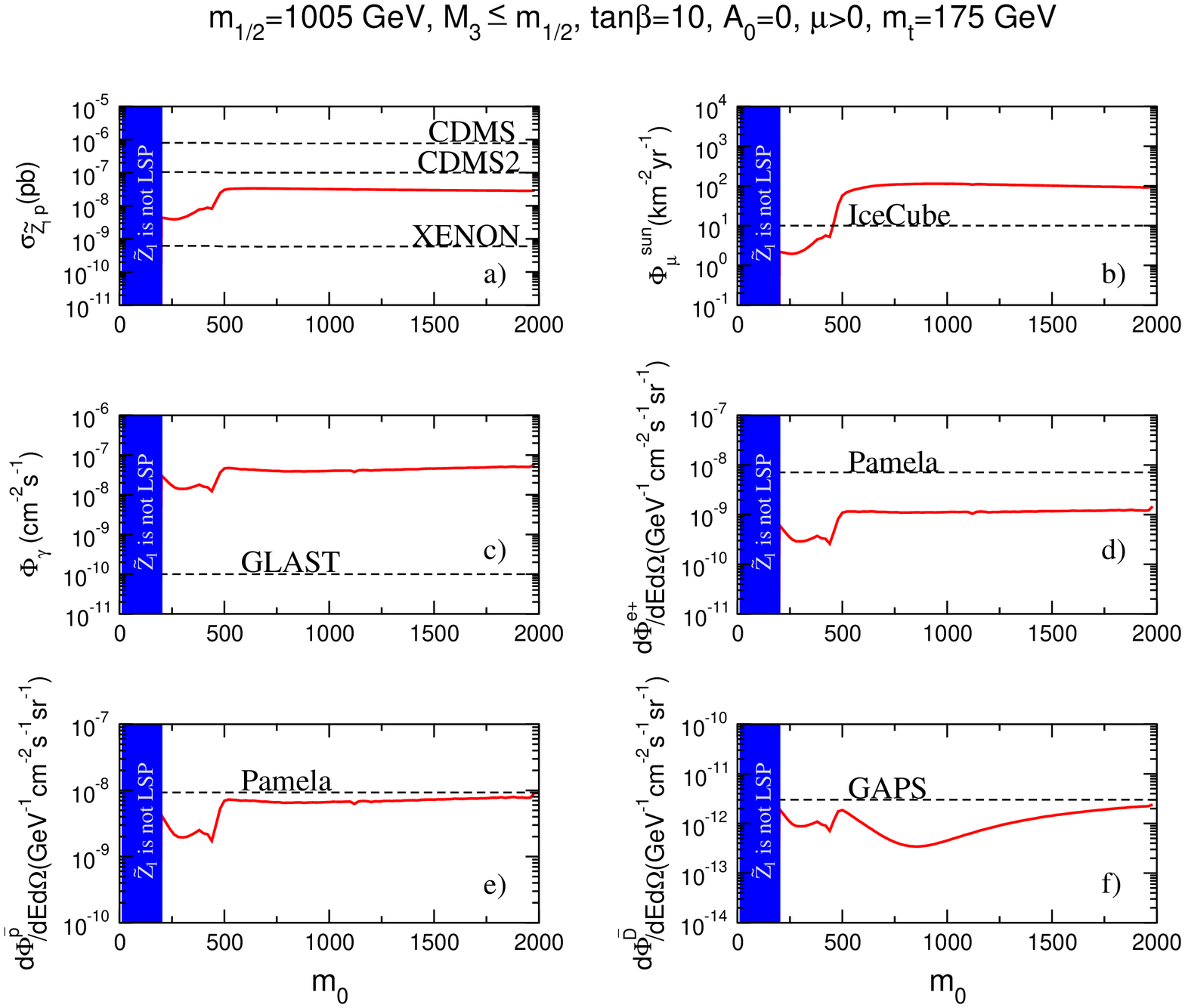,width=12cm,angle=-90} }
\caption{\label{fig:dmdet1005}
Rates for direct and indirect detection of neutralino dark matter
vs. $m_0$ for the $m_{1/2}=1005$~GeV slice of the plane in
Fig.~\ref{fig:r3}{\it b} together with the projected sensitivities of
the various experimments, assuming the adiabatically contracted N03
distribution for halo dark matter.}}

We present an overview of the reach of direct and indirect dark matter
search techniques in Fig. \ref{fig:dmdet_dirnt} and
Fig. \ref{fig:dmdet_am}. In both figures, we show the 
boundary of the region where there will be an observable signal
in the corresponding dark matter detection channel: the signal {\em  will
be detectable} in the direction indicated by the arrows. 
In Fig.~\ref{fig:dmdet_dirnt} we show results for those detection
techniques for which the reach is mainly determined by 
the local dark matter density and average
circular velocity at the Sun location (here assumed to respectively be
$\rho_{DM}(r_0)=0.38\ {\rm GeV\ cm}^{-3}$ and $\overline{v}(r_0)=221\
{\rm km\ s}^{-1}$) and is relatively less sensitive to the {\em dark matter
halo} profile, namely direct detection and the flux of energetic
neutrinos from the center of the Sun originated by neutralino
annihilations.  In Fig. \ref{fig:dmdet_am} we collect, instead, those
quantities whose dependence on the details of the dark matter halo is
more critical, namely antimatter fluxes from neutralino annihilations in
the galactic halo. For the latter, we adopt the Adiabatically Contracted
N03 Halo model\cite{n03}. We do not show the sensitivity contours for
GLAST, since with this choice for the dark matter halo profile all of the
parameter space is within the reach of the space-borne gamma-ray
detector.

We notice that while the prospects for {\em Stage-2} detectors do not
look particularly promising in this context, with the possible exception
of small regions at low neutralino masses and with light scalars, next
generation {\em Stage-3} detectors will have a sensitivity which we
estimate to be able to cover most of the parameter space of the models
under consideration. The sensitivity of IceCube will be instead
critically dependent on the higgsino fraction of the lightest
neutralino, which not only controls the pair annihilation cross section,
but, more critically, sets the spin-dependent neutralino-nucleon
scattering cross section. The reach contour we find follows in fact
quite closely the LSP higgsino fraction. However, close to the boundary
of the LEP excluded region at the bottom of frames {\it a}) and {\it
b}), the neutralino mass is so low that
the annihilation neutrinos are not energetic enough
to give the required muon flux above the IceCube detection threshold.

Turning to antimatter searches, 
we assess here the Pamela sensitivity for primary antiproton and positron
fluxes following the approach of Ref. \cite{antimatter}, evaluating the
projected total $\chi^2$ and demanding a statistically significant (at
95\% C.L.) excess over the estimated 
background.\footnote{Specifically, 
we evaluate the Pamela sensitivity on the $(m_0,m_{1/2})$ 
planes using the
approach of Ref. \cite{antimatter}, where the authors evaluate a
prospective $\chi^2$ taking into account a background independently computed
with the Galprop package \cite{Strong:2001gh} and an estimated energy binning \cite{Picozza:2002em}. This approach is
more accurate than looking at a single energy bin, although it has been 
checked that the two approaches are in reasonable agreement.}
In Fig.~\ref{fig:dmdet_am},
we shade in grey those regions which
are already statistically excluded at the 95\% C.L. by 
current data on the antiproton fluxes, when combined with an
independently estimated secondary and tertiary background
\cite{antimatter}. We stress though that this exclusion is very sensitive to
the assumed halo profile, so is by no means a rigorous bound.
In general, we find the most promising antimatter
search technique will be the antiproton channel; remarkably
enough, within this scenario we expect a signal at space-based
antiprotons searches for neutralino masses as large as 0.5 TeV. A
low-energy antideuteron signal is also expected at GAPS even on a
balloon-borne experiment in a quite large portion of the parameter
space. Finally, a significant positron signal can be only marginally
reconciled with current constraints from antiprotons, and might take
place in the low scalar masses portion of the planes under scrutiny
here.

We notice that in general the anti-particle sensitivity contours we
obtain trace the higgsino fraction of the LSP: for very small values of
$m_0$, the sensitivity drops at lower neutralino masses even though
the antimatter primary fluxes approximately scale as $\langle\sigma
v\rangle/ m_{\tz_1}^{2}$, because the LSP becomes more bino-like. Notice
also the corridor around $m_{1/2}=400$--450~GeV where the current
anti-proton searches do not exclude the
model. For smaller values of $m_{1/2}$, the LSP mass is small and the
anti-proton flux is too large. This flux falls below the 95\% CL limit
until the LSP mass becomes large enough so that
annihilation to $t\bar{t}$ becomes kinematically allowed, once again
yielding a large anti-proton rate. Finally, when $m_{1/2}$ gets larger, the LSP mass increases, and the anti-particle rate once again drops below the current
experimental limit.

We stress that the exclusion limits shown in frames {\it c})
and {\it d}) as well as the projections for the reach via anti-particle
and gamma ray searches {\em are sensitive to our assumption of the
adiabatically contracted N03 halo profile} which yields considerably
higher values for anti-particle, and especially, gamma ray
fluxes. Assuming different but equally viable galactic DM halo
distributions \cite{bo,bbko,antimatter,nuhm} modifies this
conclusion. Until the halo profile can be independently determined, we
believe that exploration of independent signals even in these
``halo-profile-dependent excluded regions'' should continue.


%
\FIGURE[!t]{
\mbox{\hspace*{-1cm}\epsfig{file=neg_dm_dirnt.eps,width=8cm}\qquad
\epsfig{file=pos_dm_dirnt.eps,width=8cm}}
\caption{\label{fig:dmdet_dirnt} Projections for the reach 
of Stage-2
and Stage-3 direct detection experiments,  and of IceCube
in the
$(m_0,m_{1/2})$ planes of the LM3DM model with 
$r_3<0$ (left) and $r_3>0$ (right). We shade in red regions
where the stau is  lighter than the LSP, where electromagnetic
gauge invariance
is spontaneously broken, or where the lightest chargino mass is not
compatible with the LEP-II bound. The arrows denote the regions where
the signal should be detectable. }}
\FIGURE[!t]{
\mbox{\hspace*{-1cm}\epsfig{file=neg_dm_am.eps,width=8cm} \qquad
\epsfig{file=pos_dm_am.eps,width=8cm} }
\caption{\label{fig:dmdet_am} The same as Fig.~\ref{fig:dmdet_dirnt},
but for antimatter searches at PAMELA and GAPS. GLAST will have the
sensitivity to probe the entire plane.  These projections are sensitive
to our choice of the adiabatically contracted N03 halo model for the
distribution of the galactic dark matter.  The
regions shaded in grey are excluded by current antiproton data, again
assuming the adiabatically contracted N03 halo model, but not
necessarily for other equally viable halo profiles. }}

\section{LM3DM at colliders}
\label{sec:col}

\subsection{Fermilab Tevatron}
\label{ssec:tev}

In SUSY models with gaugino mass unification, the ratio of weak scale
gaugino masses is typically found to be $M_1:M_2:M_3\sim 1:2:7$.  In the
mSUGRA model, since $m_{\tw_1}\sim M_2$ and $m_{\tg}\sim M_3$, the bound
on chargino masses $m_{\tw_1}> 103.5$ GeV from LEP2 implies as well that
$m_{\tg}\agt 350$ GeV. Since a 400 GeV gluino is typically beyond the
reach of Fermilab Tevatron experiments, this leaves a relatively tiny
window for a gluino discovery at the Tevatron, at least within the mSUGRA
framework\cite{tevreach}.\footnote{For $m_{\tg}=400$ GeV and
$m_{\tq}=2m_{\tg}$, we find $\sigma (p\bar{p}\to \tg\tg ) =27.8$ fb at
$\sqrt{s}=2$ TeV.}

However, in the case of LM3DM, the gluino mass can be much reduced 
relative to the value of $m_{\tw_1}$. The situation is illustrated
in Fig. \ref{fig:mgl}, where we plot contours of $m_{\tg}$ in the 
same $m_0\ vs.\ m_{1/2}$ plane as in Fig. \ref{fig:r3} where $r_3$ has
been dialed to low enough values that the value
of $\Omega_{\tz_1}h^2 =0.11$ everywhere. Of importance here is
that the gluino mass immediately adjacent to the 
blue-shaded LEP2 excluded region reaches  values below 200 GeV,
which is surely accessible to Fermilab Tevatron searches. In fact, for 
$m_0 =1500$ GeV, $m_{1/2}=250$ GeV, we find the LM3DM spectrum labelled as 
LM3DM3 in Table \ref{tab:m3dm}, where $m_{\tg}=183$ GeV is consistent
with LEP lower limits on the chargino mass.
In this scenario, the cross section for $p\bar{p}\to \tg\tg$ at the
Fermilab Tevatron with $\sqrt{s}=2$ TeV is $\sim 21.5$ pb, so that for
1 fb$^{-1}$ of integrated luminosity, we already expect in excess of $ 20$K
$\tg\tg$ events! Thus, considerable portions of the $m_0\ vs.\ m_{1/2}$
parameter plane {\em that could not be probed at LEP}  in the LM3DM scenario 
should be accessible to present day Tevatron SUSY searches!
\FIGURE[!t]{
\epsfig{file=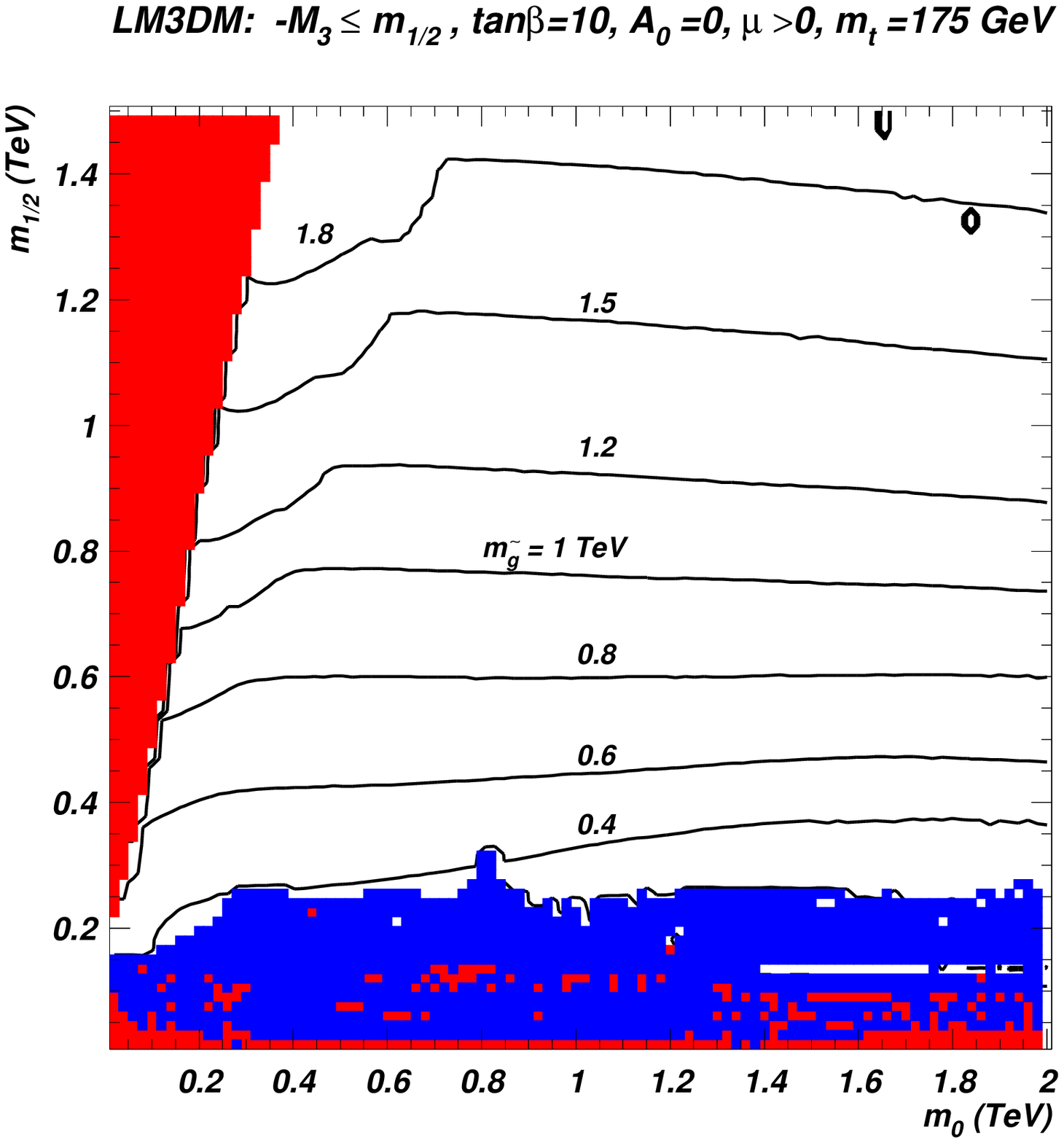,width=7cm} 
\epsfig{file=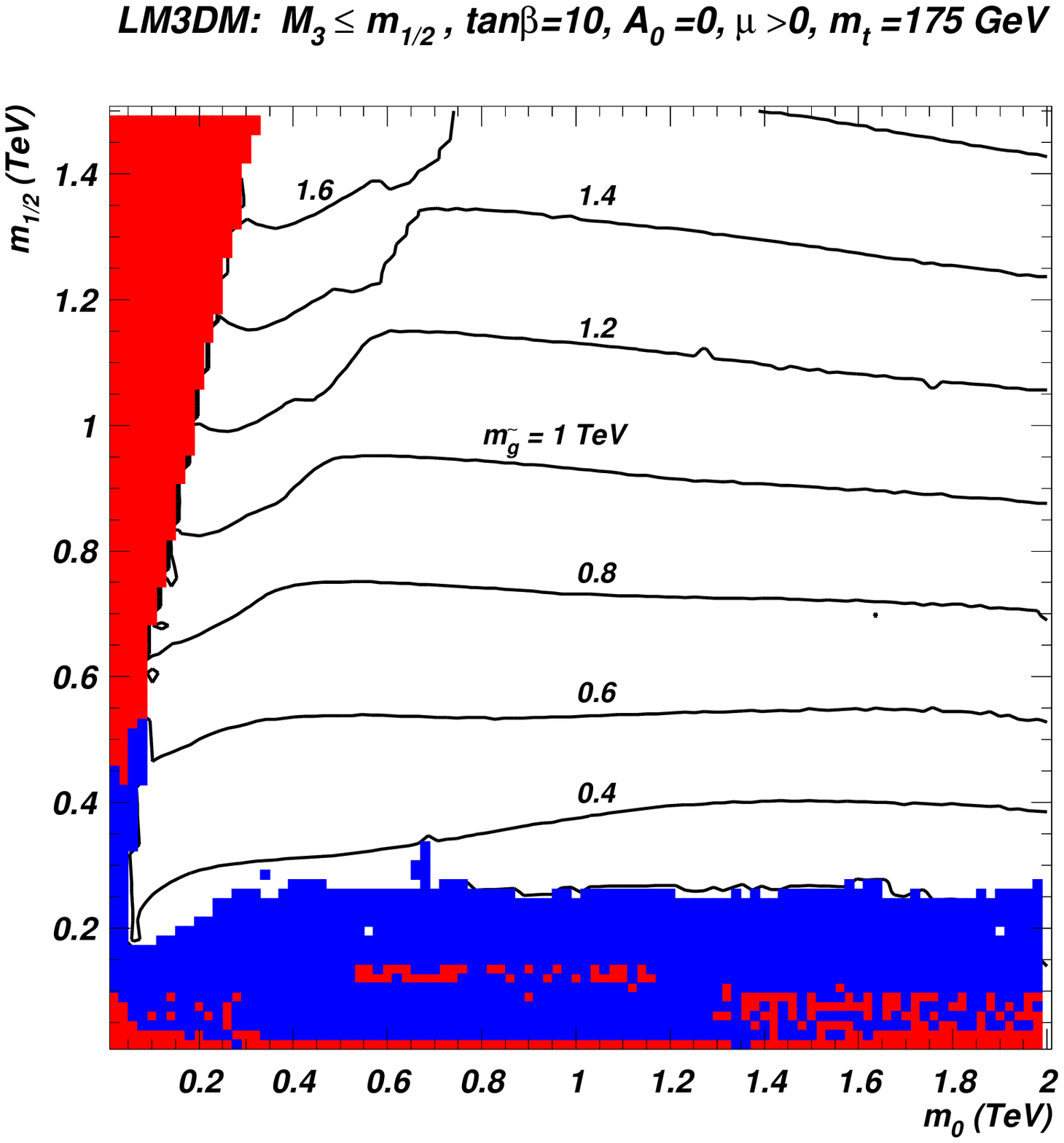,width=7cm} 
\caption{\label{fig:mgl}
Contours of $m_{\tg}$ in the $m_0\ vs.\ m_{1/2}$ plane for
$\tan\beta =10$, $A_0=0$, $\mu >0$ and
{\it a}) mSUGRA model, {\it b}) $M_3<0$ LM3DM  and
{\it c}) $M_3>0$ LM3DM.}}

Once the gluino pairs are produced, it is important to examine their
decay modes. Inspection of the Isajet decay table for point LM3DM3 shows
the surprising result that in this region of parameter space, gluino
decays via three body modes $\tg\to q\bar{q}\tz_i$ and $\tg\to q\bar{q}'
\tw_j$ are suppressed, and that in fact the loop decays $\tg\to \tz_i g$
are dominant! This large enhancement of the radiative decay relative to
three body decays is not hard to understand. It has long been
known\cite{loop} that, because of the tracelessness of the diagonal
generators of the electroweak gauge group, for degenerate squarks, {\em
all} contributions from the {\em gaugino components} of the neutralino
cancel in the amplitude for the decay $\tg \to g\tz_i$. As a result,
unless $\tan\beta$ is very large, top squark loops completely dominate
the radiative decay amplitude. The usual three-body decays of gluinos, on the
other hand, receive significant contributions from the gaugino
components of neutralinos, and indeed dominate these decays unless
$\tan\beta$ is very large. For very heavy squarks, neglecting bottom
quark Yukawa couplings and phase space effects, we  find that

\begin{equation}
\frac{\Gamma(\tg \to g\tz_i)}{\Gamma(\tg \to q\bar{q}\tz_i)}
= \frac{12}{\pi}\frac{\alpha_s f_t^2}{\left|A^q_{\tz_i}\right|^2 +
\left|B^q_{\tz_i}\right|^2} |v_1^{(i)}|^2 \left(\frac{m_t}{m_{\tg}}\right)^2
\left(\frac{m_{\tq}}{m_{\tst}}\right)^4 \left(\ln
\frac{m_{\tst}^2}{m_{\tq}^2}  - 1\right)^2\;,
\label{brfrac}
\end{equation}
where, in the notation of Ref. \cite{wss}, $A^q_{\tz_i}$ and
$B^q_{\tz_i}$ are couplings of the  $i^{th}$
neutralino (gaugino components) to the quark-squark system, and $v_1^{(i)}$ is the component
of the higgsino field $\th_u$ (that couples to up type (s)fermions) in this
neutralino. In deriving (\ref{brfrac}), we have assumed that the 
$v_1^{(i)2} \ll max(v_3^{(i)2}, v_4^{(i)2})$, {\it i.e.} that the
neutralino is mainly gaugino-like. In many models, where light
neutralinos have only small higgsino components,
gluino radiative decays have very small branching ratios because of the 
factor $v_1^{(i)2}$ in (\ref{brfrac}). This same factor is, however,
precisely the reason for the large branching fraction in the LM3DM3
scenario that we have been discussing. 

To bring home this point, we show in Fig. \ref{fig:glbf} the $\tg\to \tz_i g$
branching fraction contours summed over $i=1-4$ in the $m_0\ vs.\
m_{1/2}$ plane of Fig. \ref{fig:r3}. We see in the lower right region
where gluino masses are quite light that the cumulative gluino loop
decay branching exceeds 80\%! Thus, in this region, $\tg\tg$ production
events with $\tg\to\tz_1 g$ will give rise typically to dijet$+\eslt$
events, much like squark pair production when
$m_{\tq}<m_{\tg}$. However, the $\tg\to\tz_i g$ decays are not only into
$g\tz_1$, but also have large rates into $\tz_2 g$ and $\tz_3 g$, as
shown in Fig. \ref{fig:bf}, where these branching fractions are plotted
versus $m_0$ for fixed $m_{1/2}=300$ GeV and other parameters as in
Fig. \ref{fig:mgl}.
\FIGURE[!t]{
\epsfig{file=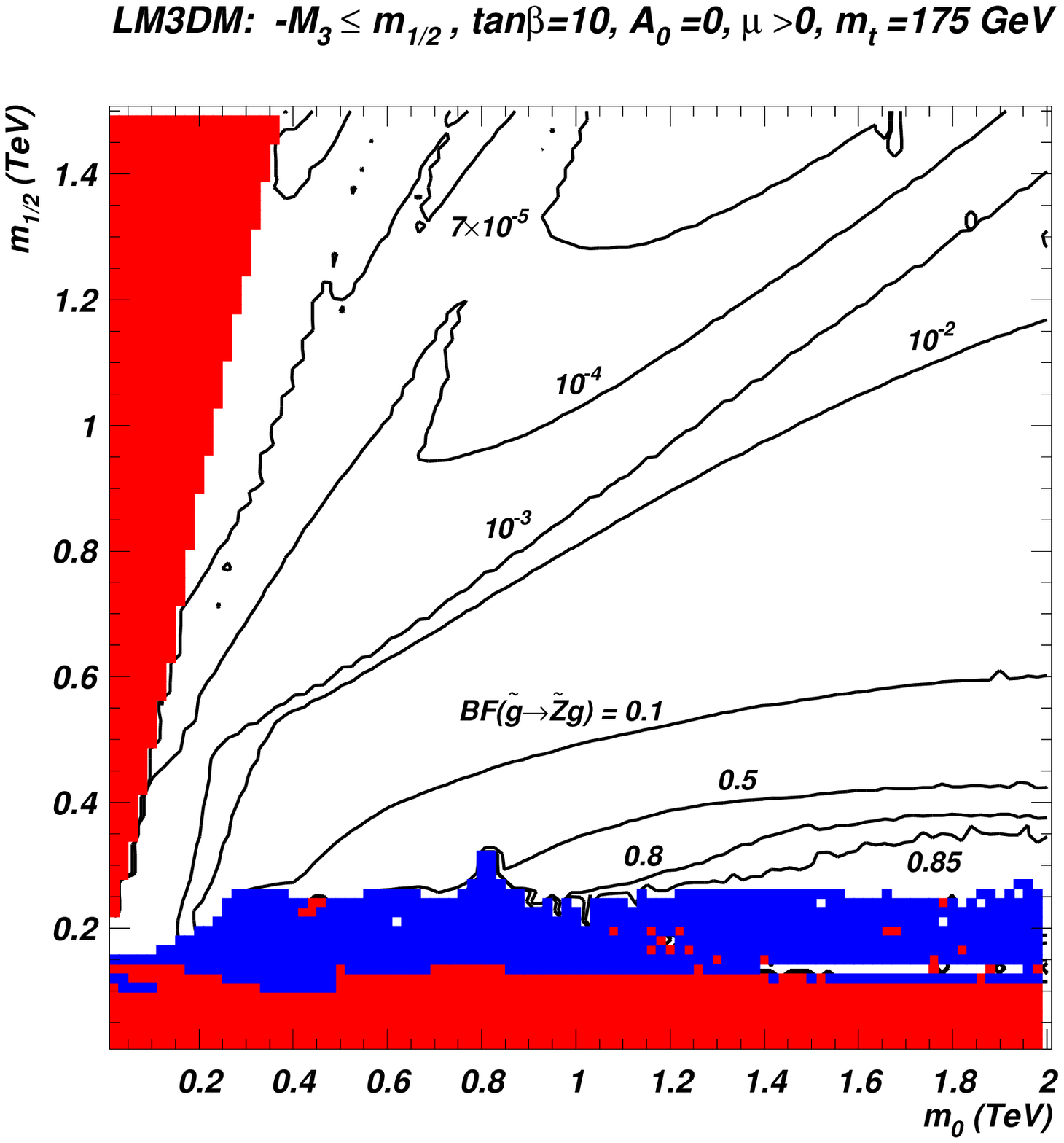,width=7cm} 
\epsfig{file=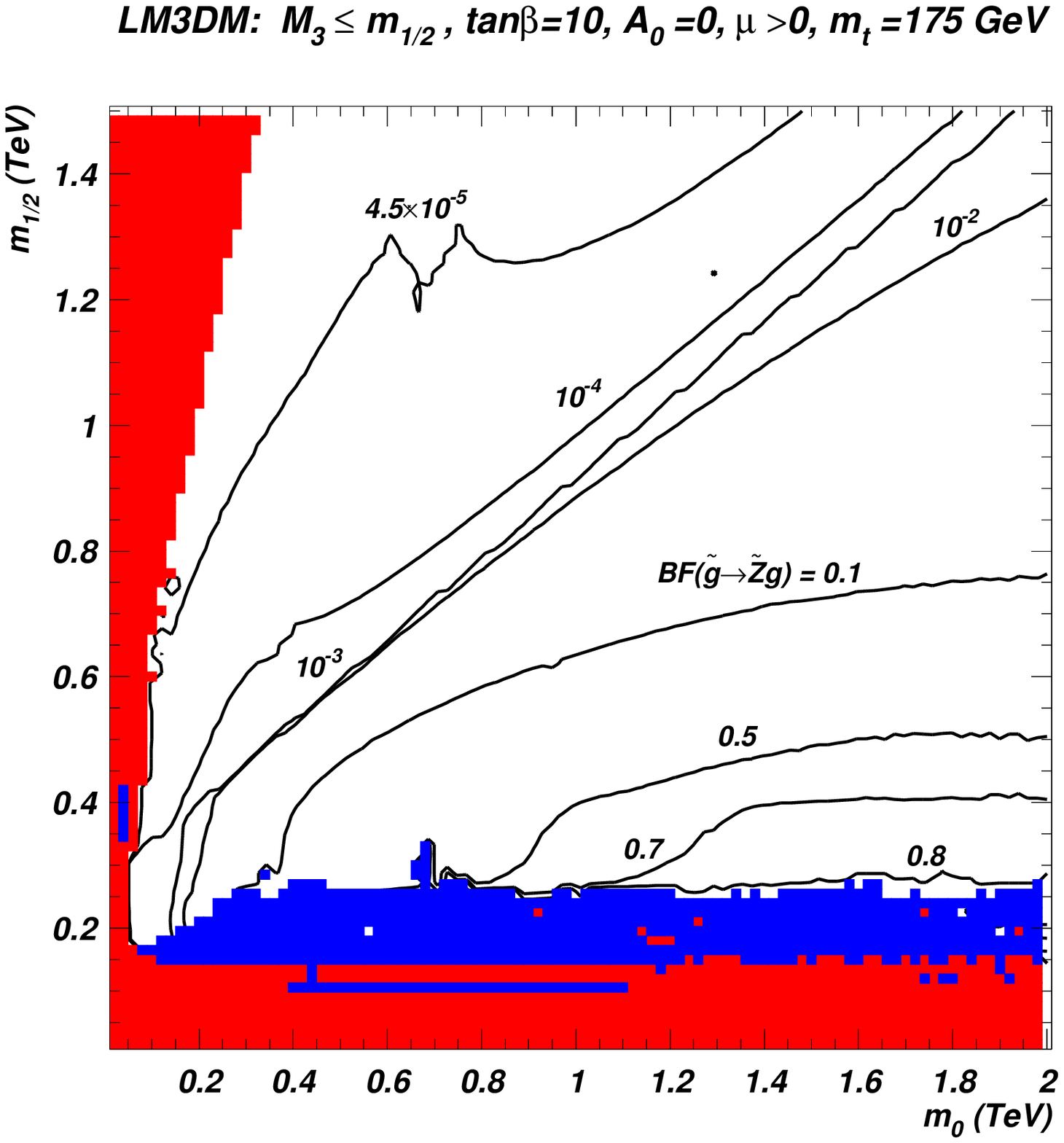,width=7cm} 
\caption{\label{fig:glbf}
Contours of $BF(\tg\to \tz_i g )$ (summed over $i=1-4$) in 
the $m_0\ vs.\ m_{1/2}$ plane for
$\tan\beta =10$, $A_0=0$, $\mu >0$ and
{\it a}) $M_3<0$ LM3DM  and {\it b}) $M_3>0$ LM3DM.} }
\FIGURE[!t]{
\epsfig{file=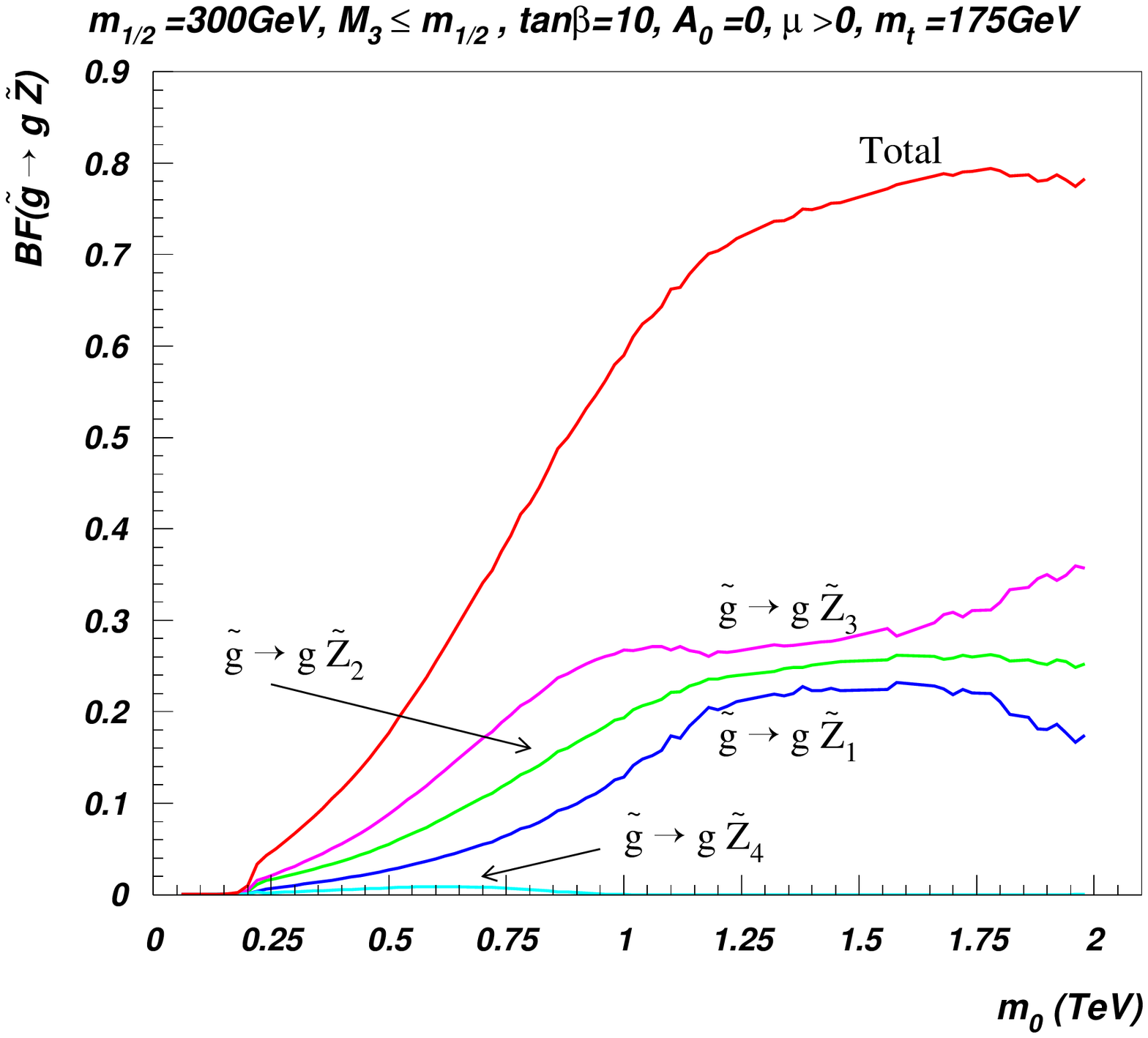,width=10cm} 
\caption{\label{fig:bf}
A plot of $BF(\tg\to \tz_i g )$ (for $i=1-4$)  
versus $m_0$ for $m_{1/2}=300$ GeV, $A_0=0$, $\tan\beta =10$, $\mu >0$
where at each point $M_3$ has been dialed so that $\Omega_{\tz_1}h^2 =0.11$}}

The $\tz_2$ and $\tz_3$ which are produced either directly 
or via gluino cascade decays will likely decay via three-body 
modes which, if $m_{\tq}$ is large enough, are dominated by $Z$
exchange. The $m_{\tz_2}-m_{\tz_1}$ mass gap is shown in Fig. \ref{z21gap}.
Since $|\mu |$ is typically quite low, and the lighter $\tz_i$
are mixed higgsino states, this mass gap varies in the 30-70 GeV
range when $\tz_1$ is MHDM, so that $\tz_2\to\ell\bar{\ell}\tz_1$
(and also frequently $\tz_3\to \ell\bar{\ell}\tz_1$) occur all over the
MHDM portion of the LM3DM parameter space. In this case, one or 
possibly even two distinct $m(\ell^+\ell^- )$ mass edges should be apparent
if enough sparticle pair production events are generated. 
The $m(\ell^+\ell^- )$ mass edges of course are renown for being 
the starting point for sparticle mass reconstruction at 
hadron colliders\cite{mlledge}.

\FIGURE[!t]{
\epsfig{file=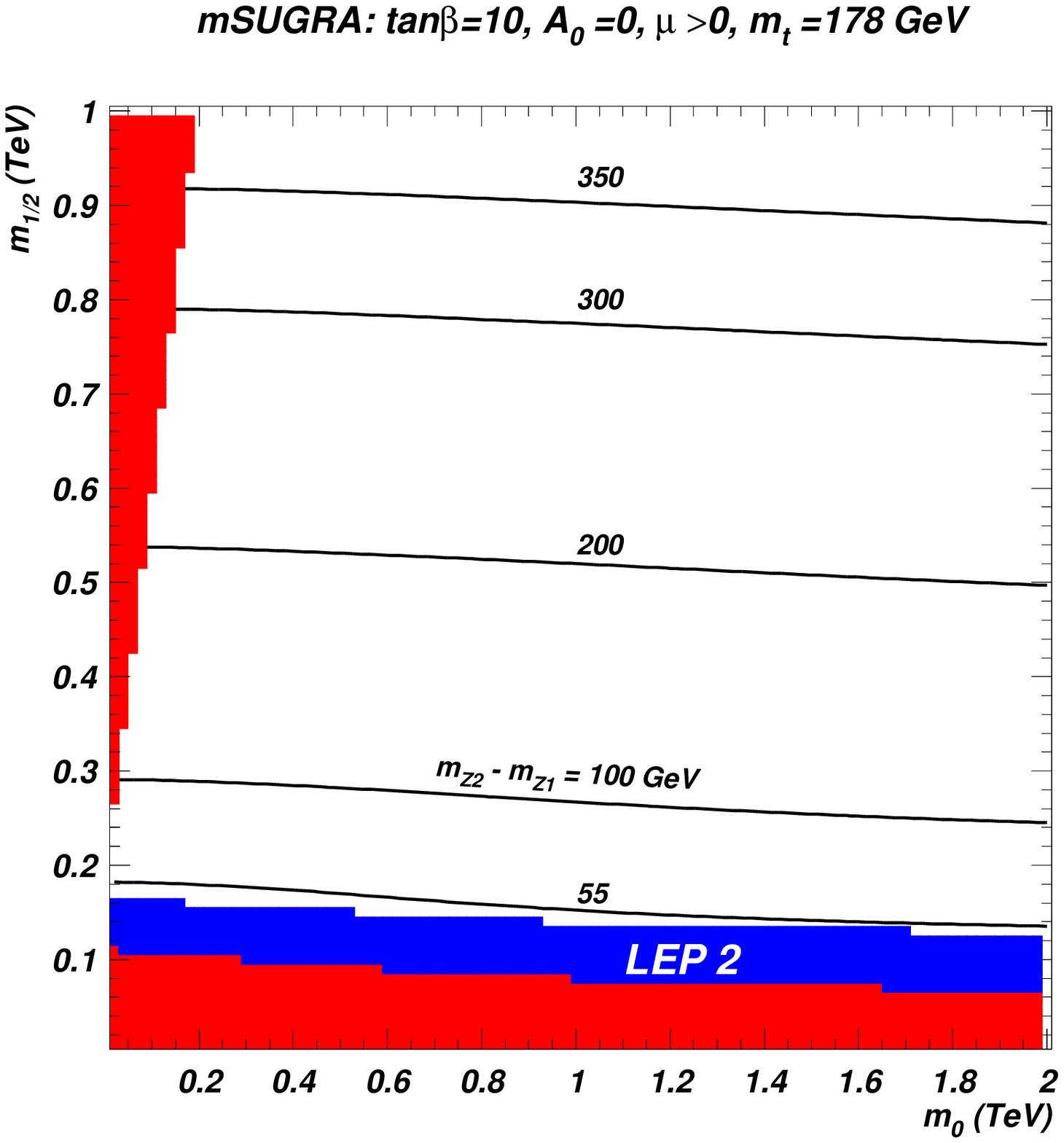,width=7.5cm} 
\mbox{\hspace*{-1cm}\epsfig{file=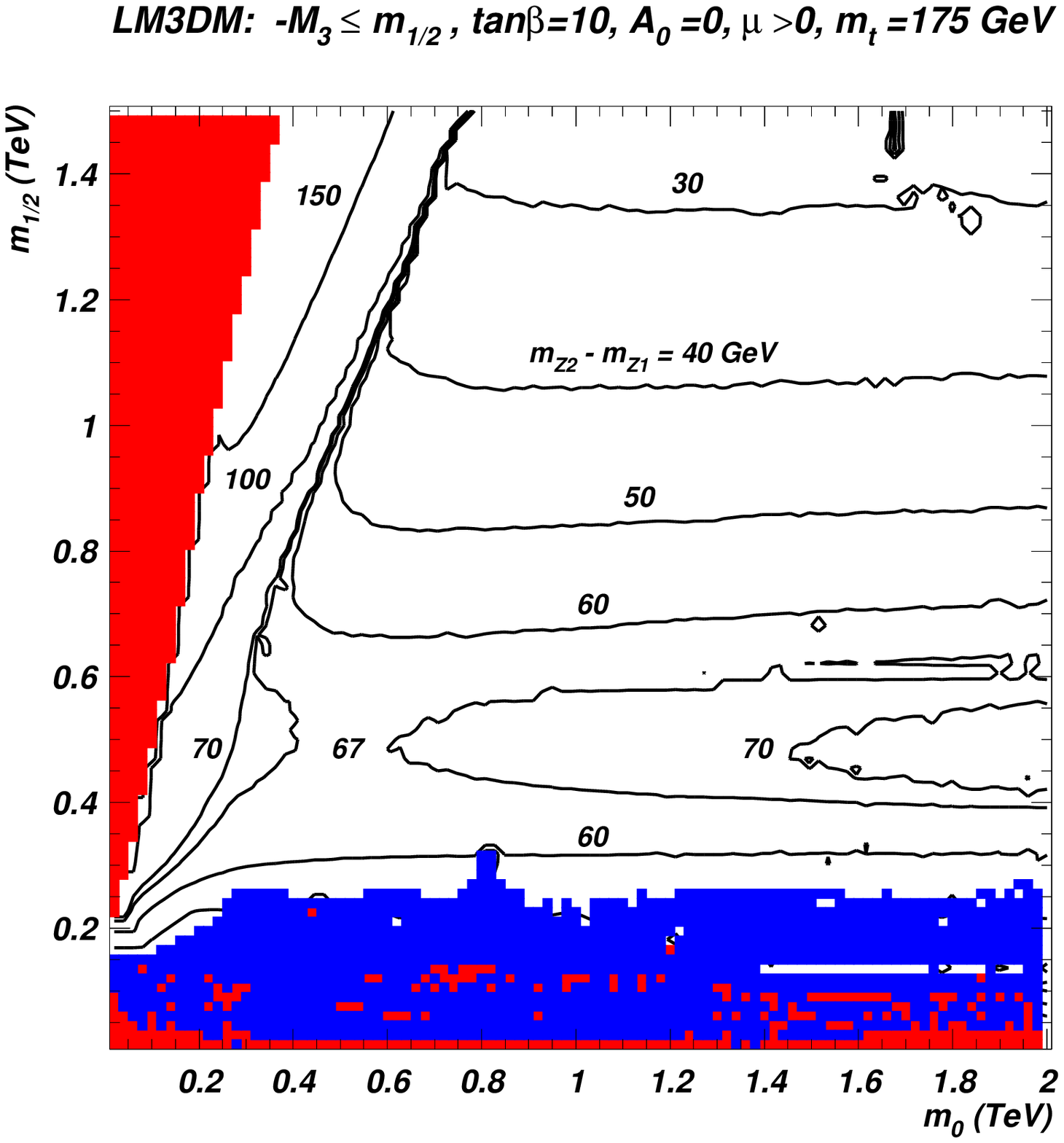,width=7.5cm} 
\epsfig{file=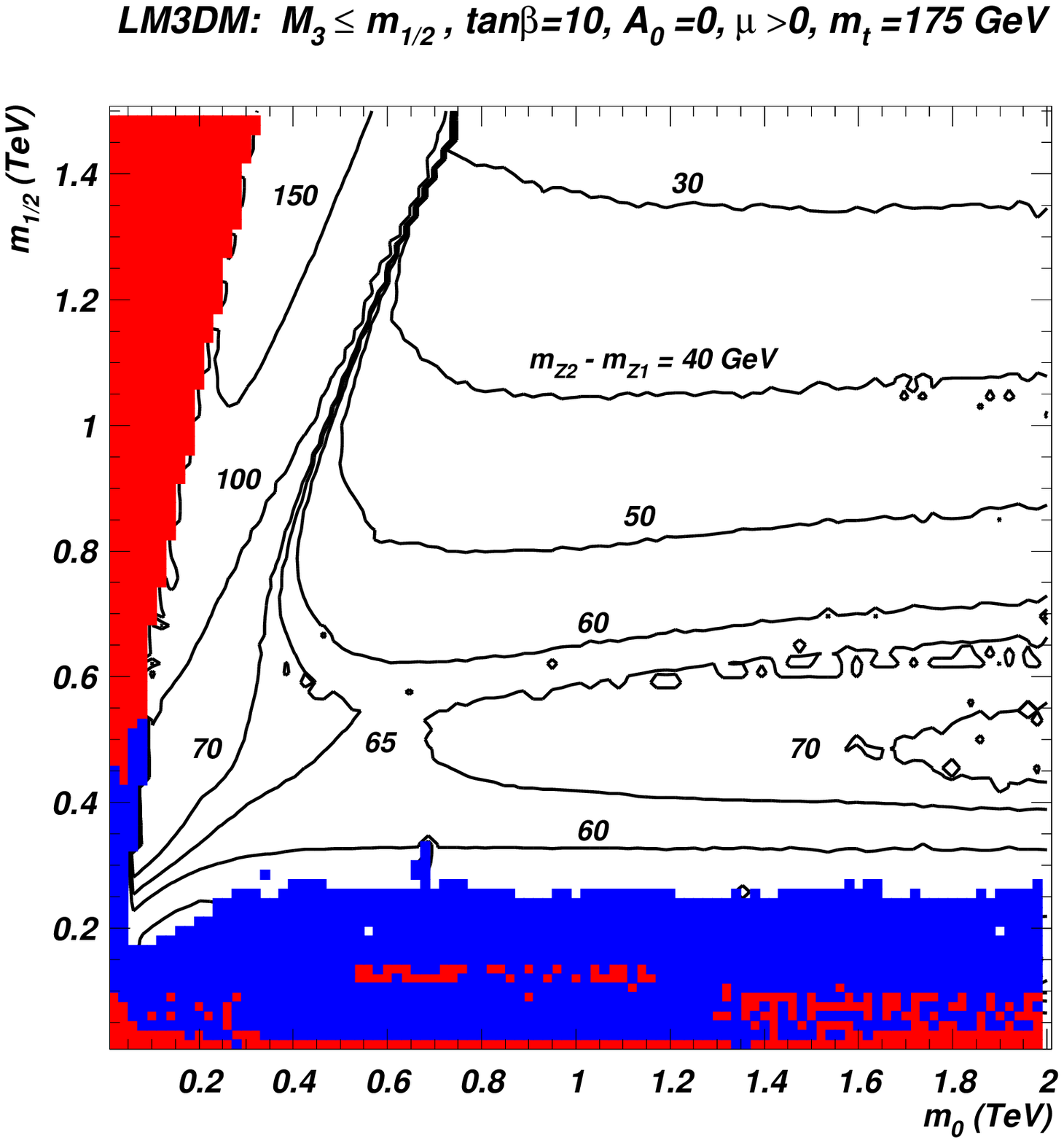,width=7.5cm} }
\caption{\label{z21gap}
Contours of $m_{\tz_2}-m_{\tz_1}$ mass gap in 
the $m_0\ vs.\ m_{1/2}$ plane for
$\tan\beta =10$, $A_0=0$, $\mu >0$ in the case of
{\it a}) the mSUGRA model, {\it b}) $M_3<0$ LM3DM  and
{\it c}) $M_3>0$ LM3DM.}}

\subsection{CERN LHC}
\label{ssec:lhc}

When comparing the LM3DM scenario to the mSUGRA model, we have found
that for a given point in the $m_0\ vs.\ m_{1/2}$ plane, gluino and
squark masses are quite suppressed relative to slepton masses and
somewhat suppressed relative to chargino masses. The upshot is that
sparticle production at the LHC should be even more dominated by gluino
and squark production (compared to other sparticle production reactions)
than in the case of mSUGRA.  Indeed, for the points listed in Table
\ref{tab:m3dm} we find $\sigma (pp\to \tg\tg X )=2,\ 84,\ 31$ and $4760$
pb , for mSUGRA, LM3DM1, LM3DM2 and LM3DM3, respectively.  The
signatures from gluino and squark pair production at the LHC from LM3DM
will consist of various multi-jet plus multi-lepton plus $\eslt$ events
as in mSUGRA\cite{susycascade}.  However, in the lower right portion of
the $m_0\ vs.\ m_{1/2}$ plane, the SUSY events should consist mainly of
dijet $+\eslt$ events when $\tg\to \tz_1 g$, with additional jets and
opposite sign/same flavor (OS/SF) dilepton events coming from
$\tg\to\tz_2 g$ or $\tz_3 g$ followed by $\tz_2$ or $\tz_3$ decay.  In
particular, same sign dileptons, which are somewhat characteristic of
gluino pair production\cite{ssdilep}, will be relatively suppressed when
the gluino loop decays are dominant.

The LHC reach for SUSY in the mSUGRA model has been calculated in
Ref. \cite{susylhc,bbbkt} assuming 100 fb$^{-1}$ of integrated
luminosity.  The ultimate reach of the LHC is mainly dependent on the
value of $m_{\tg}$ and $m_{\tq}$, and not so dependent on their
particular decay modes. Thus, we may translate the mSUGRA reach results
into contours into the $m_{\tq}\ vs.\ m_{\tg}$ plane, and then convert
this contour into a reach contour in the $m_0\ vs. m_{1/2}$ plane of the
LM3DM scenario. The translated reach contour is shown in
Fig. \ref{fig:reach} for the same parameters as in Fig. \ref{fig:r3}{\it
b}). The reach contour on the left portion of the plot in the BDM region
mainly follows the $m_{\tg}\simeq m_{\tq}\simeq 3$ TeV contour.  The
upper left increase in reach is due to the sliver of MHDM region at low
$m_0$ and large $m_{1/2}$ shown in Fig. \ref{fig:r3}{\it b}), where
$r_3$ is relatively reduced. At intermediate $m_0$, neutralino
annihilation is assisted by the $A$ resonance, and so higher $r_3$
values are found to match the WMAP constraint (recall in this region the
$\tz_1$ is bino-like).  On the right hand side of the plot for $m_0
>1.2$ TeV, we are in the MHDM region with lower $r_3$ values, and
consequently lighter gluino and squark masses. Thus, the reach is
increased.  At very large $m_0$ values, $m_{\tq}>m_{\tg}$, and here the
LHC reach extends only out to $m_{\tg}\sim 2.7$ TeV. We should mention
that since light neutralinos and chargino have significant higgsino
component in the LM3DM scenario, gluino decays to the third generation
quarks will be enhanced exactly as in the FP/HB region of the mSUGRA
model, so that $b$-jet tagging may improve the LHC reach by $\sim
10-15$\% beyond that shown in the figure.\cite{mmt}

\FIGURE[!h]{
\epsfig{file=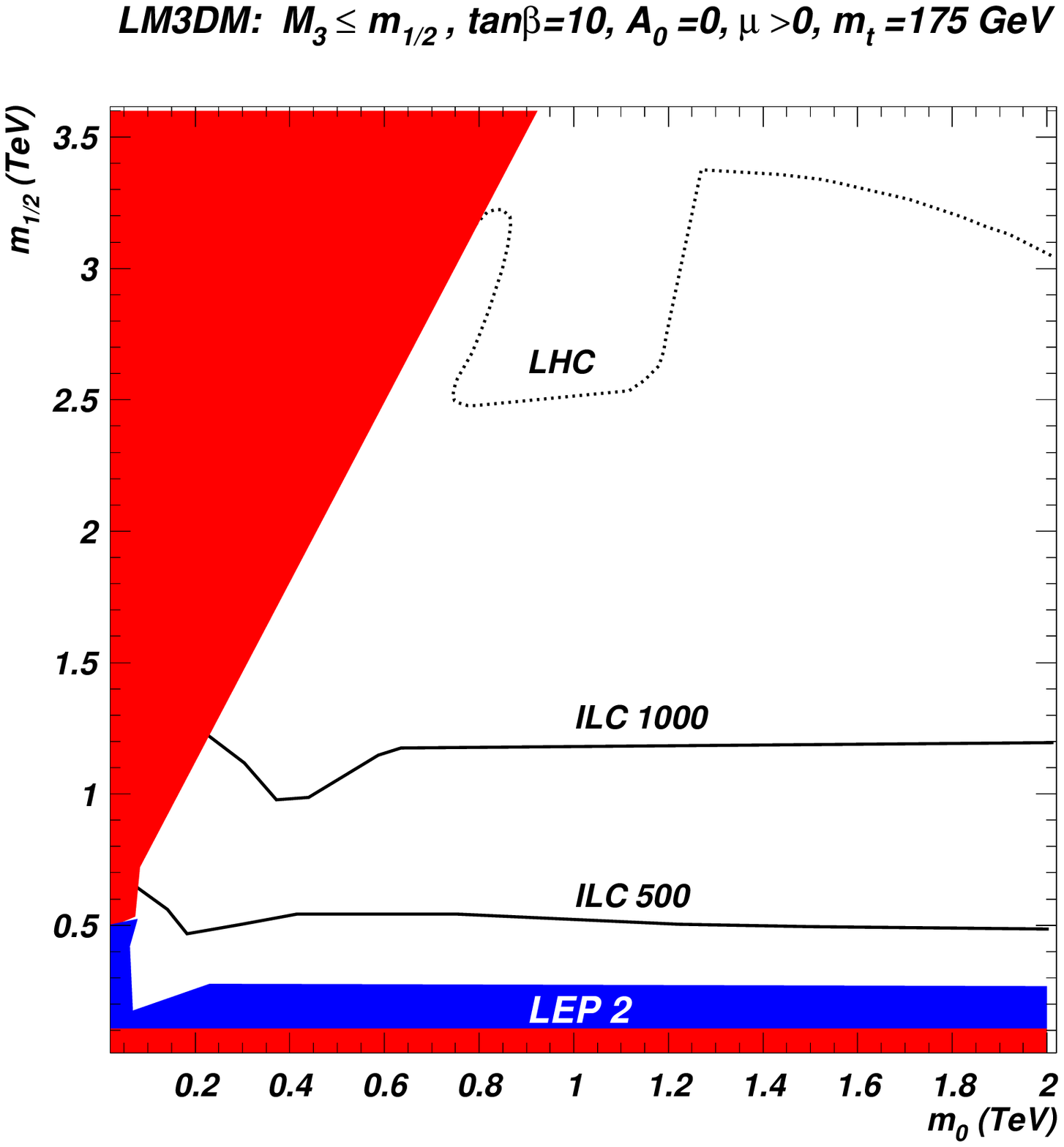,width=9cm} 
\caption{\label{fig:reach}
Reach contours for the CERN LHC with 100 fb$^{-1}$ of integrated
luminosity and for a $\sqrt{s}=0.5$ and 1 TeV linear $e^+e^-$ 
collider in 
the $m_0\ vs.\ m_{1/2}$ plane for
$\tan\beta =10$, $A_0=0$, $\mu >0$ and where $M_3$ has been reduced 
such that $\Omega_{\tz_1}h^2 =0.11$ at every point in the plane.}}

For SUSY searches at the CERN LHC, Hinchliffe {\it et al.} have 
pointed out\cite{frank} that an approximate value of $m_{\tq}$ or
$m_{\tg}$ can be gained by extracting the maximum in the
$M_{eff}$ distribution, where 
$M_{eff}=\eslt +E_T(jet\ 1)+E_T(jet\ 2)+E_T(jet\ 3)+E_T(jet\ 4)$.
Their analysis will carry over to much of the LM3DM 
scenario,\footnote{An exception is the low $m_{1/2}$, large $m_0$ region of
Fig. \ref{fig:r3} where dijet production is instead dominant.} as well as in 
models with gaugino mass unification, so that the approximate
mass scale of strongly interacting sparticles will be known soon 
after a supersymmetry signal has been established.  

In mSUGRA, a dilepton mass edge should be visible in SUSY signal events
only if $m_{1/2}\alt 250$ GeV (where $\tz_2\to\tz_1\ell\bar{\ell}$ is
allowed) or if two body $\tz_2\to \tell\bar{\ell},\ \bar{\tell}\ell$
decays are allowed.  In the case of LM3DM, as with MWDM\cite{winodm} and
BWCA DM\cite{binodm}, the dilepton mass edge should be visible over
almost all parameter space. We illustrate the situation for three of the
case studies listed in Table \ref{tab:m3dm}.\footnote{In this study, a
toy detector simulation is employed with calorimeter cell size
$\Delta\eta\times\Delta\phi=0.05\times 0.05$ and $-5<\eta<5$. The
hadronic energy resolution is taken to be $80\%/\sqrt{E}$ for
$|\eta|<2.6$ and $100\%/\sqrt{E}$ for $|\eta|>2.6$. The electromagnetic
energy resolution is assumed to be $3\%/\sqrt{E}$. We use a UA1-like jet
finding algorithm with jet cone size $R=0.5$ and $p_T^{\rm jet}>25$ GeV.
We also require that $|\eta_{\ell}|<2.5$ and $|\eta_j|<3$.  Leptons
($e$s or $\mu$s) have to also satisfy $p_T^{\rm lepton} \ge 10$~GeV.
Leptons are considered isolated if the visible activity within the cone
$\Delta R<0.3$ is $\Sigma E_T^{\rm cells}<2$ GeV. The strict isolation
criterion helps reduce multi-lepton background from heavy quark
(especially $t\bar t$) production. }  The first case, labeled mSUGRA,
has $m_0=m_{1/2}=300$ GeV, with $A_0=0$, $\tan\beta =10$ and $\mu >0$.
In this case, $\tg\tg$, $\tg\tq$ and $\tq\tq$ production occurs with a
combined cross section of about 12 pb, while the total SUSY cross
section is around 13.4 pb (the additional 1.4 pb comes mainly from -ino
pair production and -ino-squark or -ino-gluino associated production).
The case of LM3DM1, with $M_3=150$ GeV, has a total cross section of 215
pb. The case of LM3DM2, with slightly heavier squark and gluino masses,
has a total production cross section of 101 pb. The special case of
LM3DM3, which should be accessible to Tevatron searches via its light
gluino, has a total SUSY cross section at LHC of 3744 pb.

We have generated 750K 
LHC SUSY events for the cases LM3DM1 and LM3DM2 using Isajet 7.73, 
and passed them through 
a toy detector simulation as described above. 
We adopt cuts which are similar to those 
of LHC point 5 of the study of Hinchliffe {\it et al.}\cite{frank}, 
which efficiently select the SUSY signal
while essentially eliminating SM backgrounds:
$\eslt >max(100\ {\rm GeV}, 0.2 M_{eff})$,
at least four jets with $E_T>50$ GeV, where the hardest jet has
$E_T>100$ GeV, transverse sphericity $S_T>0.2$ and $M_{eff}>800$ GeV.

In these events, we require at least two isolated leptons, and then plot
the invariant mass of all same flavor/opposite sign dileptons.  The
results are shown in Fig. \ref{fig:mll}. In the case of the mSUGRA
model, frame {\it a}), there is a sharp peak at $m(\ell^+\ell^- )\sim
M_Z$, which comes from $\tz_2\to \tz_1 Z^0$ decays where $\tz_2$ is
produced in the gluino and squark cascade decays.  In the case of LM3DM1
in frame {\it b}), we clearly see a continuum distribution with a mass
edge at $m(\ell^+\ell^- )<m_{\tz_2}-m_{\tz_1}=58.5$ GeV. We also see
events beyond this edge along with a peak at $M_Z$. In this case,
$m_{\tz_3}-m_{\tz_1}$ =90.3~GeV is within $\Gamma_Z$ of $M_Z$ and we
would expect that dileptons from $\tz_3 \to \tz_1\ell\bar{\ell}$ decays
have their mass sharply peaked just below $M_Z$. This peak would also be
populated by $Z$ bosons produced via $\tz_4 \to \tz_{i} Z^0$ or $\tw_2\to
\tw_1Z^0$ decays.
The cross section 
plotted here is $\sim 188$ fb, which would correspond to 
19K events in 100 fb$^{-1}$ of integrated luminosity (the sample 
shown in the figure contains just 646 events). 
In frame {\it c})-- with a cross section of $\sim 207$ fb 
(but just 1550 actual entries)-- once again we see the $Z^0$ peak
from decays of heavier charginos and neutralinos to the $Z$ boson,
together with a mass edge at 
$m(\ell^+\ell^- )<m_{\tz_2}-m_{\tz_1}=61$ GeV, and a continuum in
between, presumably mainly from chargino pairs in SUSY events. 
In both these LM3DM cases, the $m_{\tz_2}-m_{\tz_1}$ mass edge should be
very precisely measurable. It should also be clear that this edge is
inconsistent with models based on gaugino mass unification, in that the
projected ratios $M_1:M_2:M_3$ will not be in the order $1:\sim 2:\sim
7$ as in mSUGRA. Although the $\tz_2 -\tz_1$ mass edge will be directly
measurable, the absolute neutralino and chargino masses will, as usual,
be more difficult to extract at the LHC.
\FIGURE[ht]{
\epsfig{file=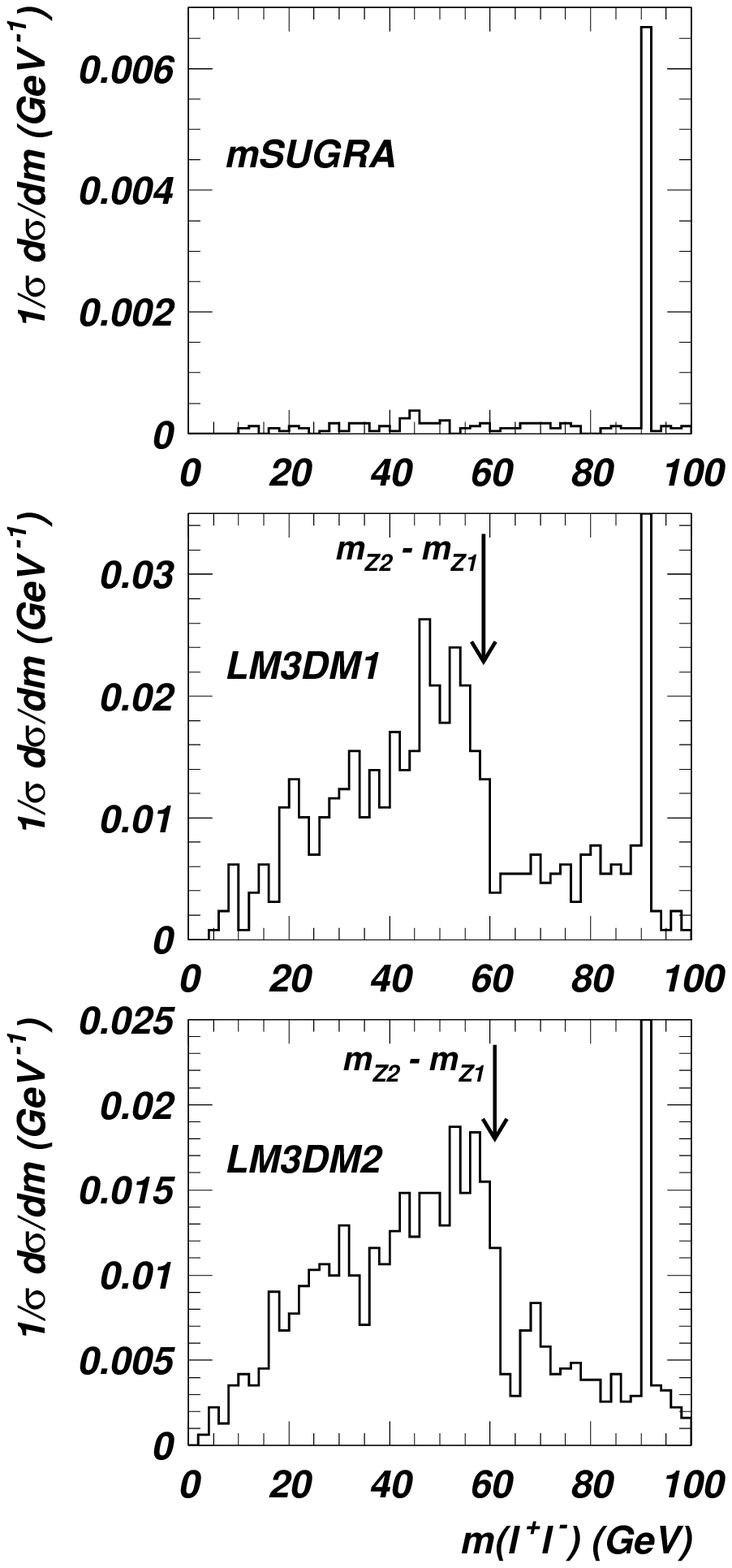,width=9cm} 
\caption{Distribution of same flavor/opposite sign dileptons from SUSY 
events at the CERN LHC 
from {\it a}) mSUGRA, {\it b}) LM3DM1 and {\it c}) LM3DM2 cases 
as in Table \ref{tab:m3dm}.
\label{fig:mll}}}

\subsection{Linear $e^+e^-$ collider}
\label{ssec:ilc}

The reach of the CERN LHC for supersymmetric matter is determined mainly
by $m_{\tq}$ and $m_{\tg}$, which depend on $m_0$ and $M_3$.  In
contrast, the reach of the ILC for SUSY is largely determined by whether
or not the reactions $e^+e^-\to \tw_1^+\tw_1^-$ or $e^+e^-\to
\tell^+\tell^-$ are kinematically accessible\cite{nlc}.  For instance,
chargino pair production is expected to be visible if
$\sqrt{s}>2m_{\tw_1}$.  The value of $m_{\tw_1}$ depends mainly on $M_2$
and $\mu$.  Thus, in the LM3DM scenario where $M_1$ and $M_2$ take
values similar to mSUGRA, but where $\mu$ is quite small, the reach of
the ILC in the $m_0\ vs.\ m_{1/2}$ plane via chargino pair production
will be enhanced relative to the case of the mSUGRA model.  Since
slepton masses are relatively unaffected by lowering $M_3$, the ILC
reach for slepton pair production will be similar to the mSUGRA case.
In addition, squark masses are relatively suppressed in the LM3DM
scenario, especially the top squark, so that there will be a non-trivial
reach of the ILC for $\tst_1\bar{\tst}_1$ production.  The situation is
illustrated in Fig. \ref{fig:ilcreach} where we show the ultimate reach
of the ILC in the $m_0\ vs.\ m_{1/2}$ plane for $\tan\beta =10$,
$A_0=0$, $\mu >0$ and $m_t=175$ GeV. We have dialed $M_3$ at every point
to give $\Omega_{\tz_1}h^2 =0.11$, in accord with the WMAP
observation. The reach of ILC with $\sqrt{s}=0.5$ TeV is denoted by
dashed contours, and extends to $m_{1/2}\sim 500$ GeV, while the
corresponding reach within the mSUGRA framework with gaugino mass
unification extends to $m_{1/2}\sim 320$ GeV\cite{nlc}.  The reach of
ILC with $\sqrt{s}=1$ TeV extends to $m_{1/2}\sim 1-1.2$ TeV, compared
with the mSUGRA value of $m_{1/2}\sim 600$ GeV.  The combined reach
(from chargino and selectron production) of the $\sqrt{s}=0.5$ and 1 TeV
ILC relative to the LHC are shown in Fig. \ref{fig:reach}. The LHC reach
is always larger than that of the ILC, primarily because of the relative
reduction of gluino and squark masses in the LM3DM framework.
%
\FIGURE[!t]{
\epsfig{file=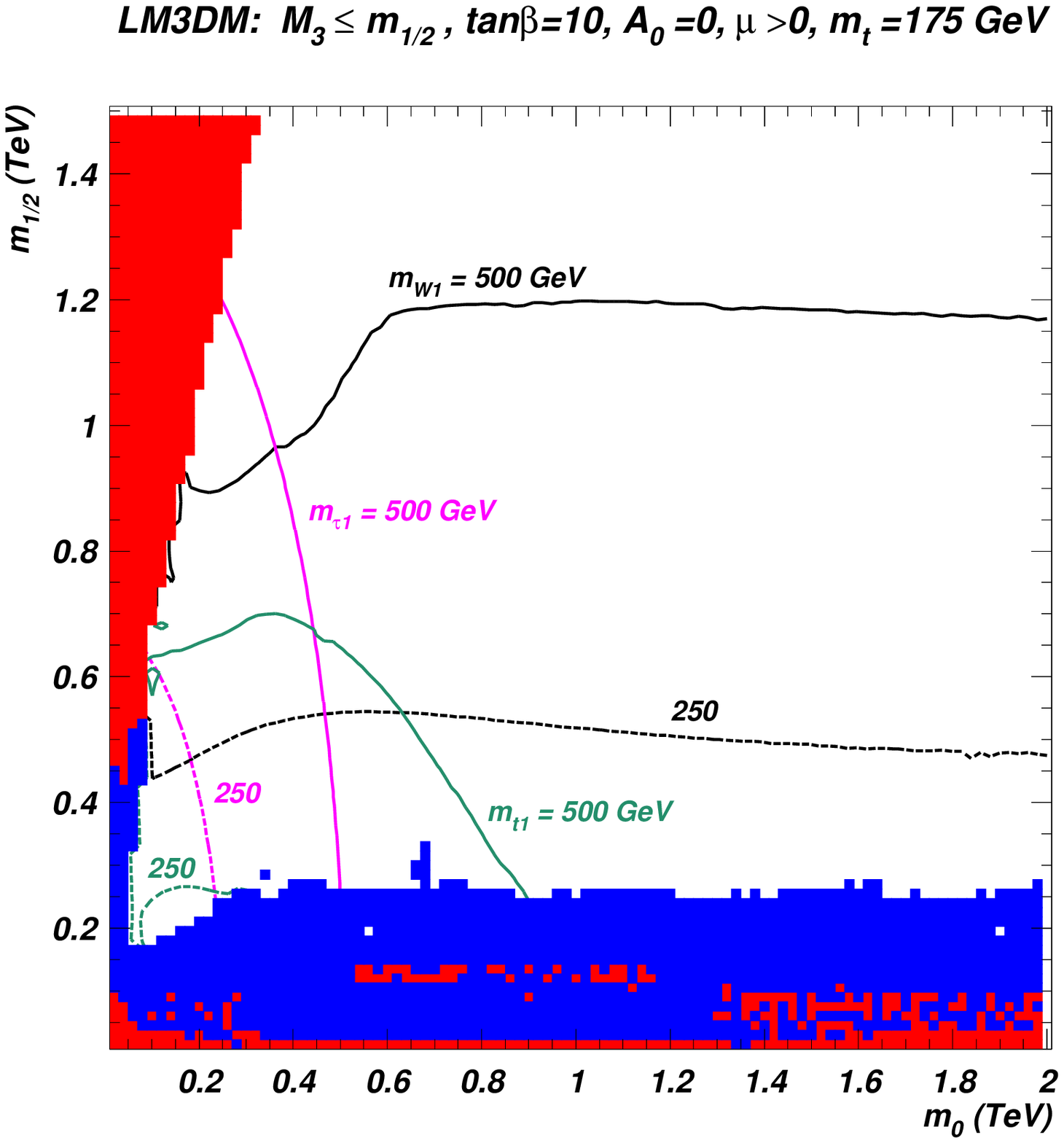,width=9cm} 
\caption{\label{fig:ilcreach}
Reach of the ILC for SUSY in the LM3DM scenario
where $M_3$ is lowered until $\Omega_{\tz_1}h^2=0.11$ at every
point in the $m_0\ vs.\ m_{1/2}$ plane. We show the ILC reach for
$\sqrt{s}=0.5$ TeV and 1 TeV via the kinematic limits for
$\tw_1^+\tw_1^-$, $\ttau^+\ttau^-$ and $\tst_1\bar{\tst}_1$
production. }}

The distinguishing feature of LM3DM is that the small $\mu$
parameter gives rise to a rather light spectrum of the two charginos and
all four neutralinos. Thus, many more -ino pair production reactions
are likely to be accessible to a linear $e^+e^-$ collider than would occur
in the mSUGRA model.
As an example, we show in Table \ref{tab:eesusy} the various
SUSY cross sections in fb which are accessible
to a $\sqrt{s}=0.5$ TeV
machine for the four case studies listed in Table \ref{tab:m3dm}.
While only $\tw_1\tw_1$, $\tz_1\tz_2$ and $\tz_2\tz_2$ production
are possible in the mSUGRA model, for the LM3DM scenarios 
all ten reactions listed are
accessible at a $\sqrt{s}=0.5$ TeV linear collider,
although some of these have very low rates. It does appear though that
every chargino and neutralino is produced via some reaction with cross
section exceeding 10~fb. Detailed studies of the chargino-neutralino
sector along the lines discussed in Ref. \cite{zerwas} should be
feasible within the LM3DM scenario. 
%
\begin{table}
\begin{tabular}{lcccc}
\hline
reaction & mSUGRA & LM3DM1 & LM3DM2 & LM3DM3 \\
\hline
$\tw_1\tw_1$ & 132.0 & 312.7 & 307.7 & 538.1 \\
$\tw_1\tw_2$ & ---   & 59.5  & 52.7  & 49.5  \\
$\tz_1\tz_2$ & 22.7  & 48.2  & 45.1  & 3.0   \\
$\tz_1\tz_3$ & ---   & 32.5  & 29.8  & 86.8   \\
$\tz_1\tz_4$ & ---   & 3.2   & 3.0   & 0.02   \\
$\tz_2\tz_2$ & 12.6  & 21.6  & 18.1  & 0.6   \\
$\tz_2\tz_3$ & ---   & 99.9  & 101.2 & 53.7   \\
$\tz_2\tz_4$ & ---   & 7.4   & 4.6   & 0.2   \\
$\tz_3\tz_3$ & ---   & 0.1   & 0.07  & 0.5   \\
$\tz_3\tz_4$ & ---   & 22.9  & 11.5  & 41.6   \\
\hline
\end{tabular}
\caption{Cross sections in fb for $e^+e^-\to SUSY$
processes at a $\sqrt{s}=0.5$ TeV linear collider, for the four
case studies listed in Table 
1.}
\label{tab:eesusy}
\end{table}

Another feature of LM3DM relevant to linear $e^+e^-$ colliders is 
that the relatively low squark masses which are expected in this scenario
means that squark pair production is more likely to be possible, especially
for a $\sqrt{s}\ge 1$ TeV machine. In most cases, the decay
$\tq\to\tg q$ is kinematically allowed, so that
gluino production might be studied in the $e^+e^-$ environment, 
since they can be produced via the cascade decays of the heavier squarks.
In this case, very precise determination of squark and gluino masses may
be possible if the end point of the energy spectrum of the primary quark
jet in the decay $\tq \to q\tg$ can be identified. 

\section{Summary and conclusions}\label{sec:conclude}

If we identify the relic density of CDM (\ref{eq:wmap}), with that of
thermally produced LSPs of an R-parity conserving SUSY model, the WMAP
measurement serves as a stringent constraint on any SUSY model. It is
then interesting to explore the ramifications of this measurement, both
for collider searches for supersymmetry, as well as for direct and
indirect searches for DM at non-accelerator facilities. It is also
necessary to explore just how robust these ramifications are to changes
in the underlying SUSY framework that do not alter the successful
prediction for the LSP relic density. In previous studies, we explored
how the WMAP CMD constraint could be satisfied if
(1)~we relax the (phenomenologically unnecessary)
assumption that the Higgs
scalar mass parameters unify with sfermion mass parameters at high
scales, and (2)~if we relax the assumption of the unification of gaugino
masses, and allow the ratio $M_1/M_2$ (which controls the composition of
the LSP) to float freely. 
In this paper, we study the implications of what we dub as the low $M_3$
DM  (LM3DM) model, which is essentially the paradigm mSUGRA framework, except
that the $SU(3)$ gaugino mass is allowed to adopt any value. Following
earlier studies \cite{belanger,mn} we find that, for $m_0$ values not
hierarchically larger than $m_{1/2}$,
the value of $|\mu|$ is reduced when the GUT scale
gluino mass parameter $|M_3| < m_{1/2}$, and MHDM or BDM solutions become
viable for essentially all values of model parameters.  

The sizeable higgsino component of MHDM implies enhanced detection
rates in on-going, planned and proposed experiments searching for DM
relative to the bino LSP case more typical in mSUGRA; see
Fig.~\ref{fig:dmdet_dirnt} and Fig.~\ref{fig:dmdet_am}.  While direct
searches at stage 2 detectors such as CDMS2 can explore only a
relatively limited portion of the parameter space, the entire parameter
space should be explorable at the proposed stage 3 detectors typified by
the SuperCDMS or 1-ton XENON experiments. Indirect searches via the
detection of hard muon neutrinos from the core of the sun should also be
possible at IceCube over much of the model parameter space. Experiments
looking for anti-particles and gamma rays from the annihilation of
neutralinos in our galactic halo should also be able to detect signals
from MHDM. These projections should be interpreted with care because
they are sensitive to the precise distribution of the DM in the galactic halo.

By comparing detection rates in direct and indirect search experiments,
it is possible to qualitatively distinguish the MHDM scenario from
scenarios where the dark matter is bino-like as in mSUGRA (either in the
bulk region or in the Higgs resonance region) or with
bino-wino coannihilation yielding the WMAP value, or mixed with the wino
\cite{binodm}. Experiments at colliders will be able to provide
additional evidence in favor of one or the other of these
scenarios. 

MHDM, on the other hand, occurs in a variety of models. It may occur for
very large values of $m_0$ in the HB/FP region of the mSUGRA model, in
non-universal Higgs mass (NUHM) models where the GUT scale Higgs mass
parameters are equal but larger than the corresponding sfermion
parameters or in more general NUHM models, or, as we have just seen, in
the LM3DM models. Distinction between these various MHDM scenarios is
only possible via examination of the properties of {\em other}
sparticles which are accessible via collider searches for SUSY. 

The main distinguishing feature of the LM3DM model is that the ratio of
coloured sparticle masses to those of colour singlet sleptons, charginos
and neutralinos is smaller in the LM3DM model than in most other
models. This clearly favours SUSY searches at hadron colliders such as
the Fermilab Tevatron or the CERN LHC vis \'a vis searches at
electron-positron colliders. For instance, while the LEP lower limit on
the chargino mass greatly restricts the potential of the Fermilab
Tevatron to discover gluinos within the mSUGRA framework (or, for that
matter, in any framework with unification of gaugino masses), {\em
Tevatron searches for gluinos may yet lead to the discovery of SUSY if
SUSY is realized as in the LM3DM model}\cite{tevgl}. 
In this case, experiments at
the LHC will have a reach much larger than that of even a TeV linear
collider. Despite this, experiments at the linear collider will play a
big role in elucidating the physics and allowing us to zero in on the
underlying scenario.  Since $|\mu|$ is comparable to the weak scale
electroweak gaugino masses, it is likely that {\it all} charginos and
neutralinos will be accessible and their properties measured at a TeV
linear collider. In this case, it will be possible to directly determine
$M_1$, $M_2$, $\mu$ and $\tan\beta$. Combining these with the
determination of $m_{\tg}$ that should be possible at the LHC, we should
be able to determine that the GUT scale gluino mass is smaller than the
corresponding electroweak gaugino masses.\footnote{It may be interesting
to examine whether this can also be concluded from the determination of
two or more dilepton mass edges from the decay of neutralinos at the
LHC.} If we are lucky, the top squark and perhaps even other squarks,
may be kinematically accessible. In this case, the gluino may also be
accessible as a decay product of the squarks, and true bottom-up
sparticle spectroscopy would be possible.

\acknowledgments

This research was supported in part by the U.S. Department of Energy
grant numbers DE-FG02-97ER41022, DE-FG03-94ER40833, DE-FG03-92-ER40701 and
FG02-05ER41361, and by NASA grant number NNG05GF69G.

%

\end{document}